\documentclass[10pt,amsmath,amssymb,twocolumn,superscriptaddress,groupedaddress,nofootinbib,aps,prd,twocolumn]{revtex4-1}

\usepackage{graphicx}  
\graphicspath{{./plots/}}
\usepackage{bm}        
\usepackage{enumitem}
\usepackage{etoolbox}

\providecommand{\texorpdfstring}[2]{#1}

\newcommand{\comment}[1]{#1}

\newcommand{\lr}[1]{ \left( #1 \right) }
\newcommand{\lrs}[1]{ \left[ #1 \right] }
\newcommand{\lrc}[1]{ \left\{ #1 \right\} }
\newcommand{\vev}[1]{ \langle \, #1 \, \rangle }
\newcommand{\cev}[1]{ \langle \langle \, #1 \, \rangle \rangle }

\newcommand{\ket}[1]{ \, | #1 \rangle }
\newcommand{\bra}[1]{ \langle #1 | \, }

\newcommand{\tr}{ {\rm Tr} \, }

\renewcommand{\Re}{ {\rm Re} \, }
\renewcommand{\Im}{ {\rm Im} \, }

\newcommand{\diag}[1]{ {\rm diag} \, \left( #1 \right) }
\newcommand{\const}{ {\rm const}}

\newcommand{\expa}[1]{ \exp{\left( #1 \right)} }

\newcommand{\hX}{\hat{X}}
\newcommand{\hP}{\hat{P}}
\newcommand{\hpsi}{\hat{\psi}}

\newcommand{\xx}{\sigma_{xx}}

\newcommand{\xp}{\sigma_{xp}}
\newcommand{\pp}{\sigma_{pp}}

\newtoggle{twocolumn}
\toggletrue{twocolumn}
\newcommand{\myeqbreak}[1]{\iftoggle{twocolumn}{#1 \nonumber \\ #1}{#1}}

\usepackage[colorlinks=true]{hyperref}

\begin{document}
\sloppy

\title{Quantum chaos, thermalization and entanglement generation in real-time simulations of the BFSS matrix model}

\author{P.~V.~Buividovich}
\email{pavel.buividovich@physik.uni-regensburg.de}
\affiliation{Institut f\"ur Theoretische Physik, Justus-Liebig-Universit\"at, 35392 Giessen, Germany}
\affiliation{Institute of Theoretical Physics, University of Regensburg,
D-93053 Germany, Regensburg, Universit\"{a}tsstrasse 31}

\author{M.~Hanada}
\email{masanori.hanada@colorado.edu}
\affiliation{Department of Physics, University of Colorado, Boulder, Colorado 80309, USA}

\author{A.~Sch\"{a}fer}
\email{andreas.schaefer@physik.uni-regensburg.de}
\affiliation{Institute of Theoretical Physics, University of Regensburg,
D-93053 Germany, Regensburg, Universit\"{a}tsstrasse 31}


\begin{abstract}
We study numerically the onset of chaos and thermalization in the Banks-Fischler-Shenker-Susskind (BFSS) matrix model with and without fermions, considering Lyapunov exponents, entanglement generation, and quasinormal ringing. We approximate the real-time dynamics in terms of the most general Gaussian density matrices with parameters which obey self-consistent equations of motion, thus extending the applicability of real-time simulations beyond the classical limit. Initial values of these Gaussian density matrices are optimized to be as close as possible to the thermal equilibrium state of the system. Thus attempting to bridge between the low-energy regime with a calculable holographic description and the classical regime at high energies, we find that quantum corrections to classical dynamics tend to decrease the Lyapunov exponents, which is essential for consistency with the Maldacena-Shenker-Stanford (MSS) bound at low temperatures. The entanglement entropy is found to exhibit an expected ``scrambling'' behavior - rapid initial growth followed by saturation. At least at high temperatures the entanglement saturation time appears to be governed by classical Lyapunov exponents. Decay of quasinormal modes is found to be characterized by the shortest time scale of all. We also find that while the bosonic matrix model becomes non-chaotic in the low-temperature regime, for the full BFSS model with fermions the leading Lyapunov exponent, entanglement saturation time, and decay rate of quasinormal modes all remain finite and non-zero down to the lowest temperatures.
\end{abstract}

\maketitle

\section{Introduction}
\label{sec:intro}

Our understanding of quantum chaos has significantly advanced in recent years due to numerous correspondences between chaotic systems and black holes. In particular, it was argued that physical systems which are holographically dual to black holes are maximally chaotic, with the Sachdev-Ye-Kitaev (SYK) model \cite{Sachdev:1006.3794,Stanford:1604.07818} and the Banks-Fischler-Shenker-Susskind (BFSS) model (supersymmetric matrix model) \cite{Susskind:hep-th/9610043,Nicolai:NPB1988,Witten:hep-th/9510135} being notable examples on the quantum field theory (QFT) side. More generally, matrix quantum mechanics provides a rather generic system for studying quantum chaos \cite{Savvidy:NPB84,Polchinski:0801.3657,Susskind:0808.2096}. Despite this progress, many questions remain open. First of all, except for the SYK model, there is no demonstration of maximal chaos from the QFT side. The mechanism which leads to fast apparent thermalization of quark-gluon plasma produced in heavy-ion collisions \cite{Nastase:hep-th/0501068,Schaefer:1012.4753,Kurkela:11:1,Romatschke:1307.2539,Chesler:1309.1439,Kurkela:15:1}, which may be related to maximally chaotic features of holographic QFT, has not been understood either. Obviously, for real QCD which should describe this process, holographic duality is not directly applicable. These problems motivate the development of numerical methods for studying quantum real-time dynamics of gauge theories \cite{Bodeker:hep-ph/9801430,Son:hep-ph/9810216,Berges:1403.4849}.

Quantum chaos can be described quantitatively in terms of the exponential growth of the  out-of-time-order correlators (OTOCs)
\begin{eqnarray}
\label{otoc_def}
 C\lr{t} = \left\langle \lrs{\hat{W}\lr{t}, \hat{V}\lr{0}}^2\right\rangle \sim \expa{2 \lambda_L t}
\end{eqnarray}
of suitable operators $\hat{W}$, $\hat{V}$ \cite{Larkin:JETP1969,Stanford:1304.6483,Maldacena:1503.01409}. In the semiclassical regime, the growth of OTOCs (\ref{otoc_def}) is governed by the leading classical Lyapunov exponent $\lambda_L^0$ of a system: $C\lr{t} \sim \expa{2 \lambda_L^0 t}$, at sufficiently large $t$.

Exponential growth of OTOCs has to be contrasted with the time dependence of the conventional time-ordered correlators $G\lr{t} = \vev{\tr\lr{\hat{W}\lr{t} \hat{V}\lr{0}}}$, which are related to dissipative transport responses. Such time-ordered correlators typically exhibit exponentially decaying oscillations characterized by complex-valued quasinormal frequencies \cite{Starinets:hep-th/0205051}, the so-called quasinormal ringing \cite{Konoplya:1102.4014}.

While in classical systems Lyapunov exponents can be arbitrarily large, a universal Maldacena-Stanford-Shenker (MSS) bound\footnote{We have set $\hbar=1$ and $k_{\rm B}=1$.}
\begin{eqnarray}
\label{mss_bound_def}
 \lambda_L \leq 2 \pi T
\end{eqnarray}
on the coefficient of exponential growth of out-of-time-order correlators (\ref{otoc_def}) can be derived in quantum theory under some mild assumptions based on analyticity properties of the OTOCs \cite{Maldacena:1503.01409}. This bound is expected to be saturated by physical systems which admit holographic dual description in terms of a black holes in weakly coupled gravity (i.e. large-$N$, strong coupling limit of holographic QFT). This could be explicitly demonstrated in the SYK model \cite{Stanford:1604.07818}, which is expected to be holographically dual to a nearly extremal black hole near zero temperature \cite{Sachdev:1006.3794,Stanford:1604.07818}.

The Banks-Fischler-Shenker-Susskind (BFSS) model \cite{Susskind:hep-th/9610043,Nicolai:NPB1988,Witten:hep-th/9510135}, obtained by reducing the $9+1$-dimensional supersymmetric Yang-Mills theory down to $0+1$ dimensions, has a significantly richer dynamics than the SYK model and also admits a well-defined dual holographic description \cite{Maldacena:hep-th/9802042} in terms of black zero brane in the type IIA superstring theory. The BFSS model is also expected to saturate the MSS bound (\ref{mss_bound_def}) in the strong-coupling regime at sufficiently low temperature; actually this is the first model in the literature which has been conjectured to be a ``fast scrambler'' \cite{Susskind:0808.2096}. While the BFSS model is known to be classically chaotic and various aspects near the classical limit have been studied \cite{Berenstein:1104.5469,Berenstein:1211.3425,Hanada:1503.05562,Hanada:1512.00019,Hanada:1602.01473,Arefeva:hep-th/9710032,Arefeva:hep-th/9804021,Axenides:1707.02878,Axenides:1712.06544,Viennot:1802.08541}, so far not much is known about its real-time dynamics in the quantum regime because of the absence of suitable first-principle methods for real-time evolution of many-body quantum systems.
Note that, for exactly solvable $O\lr{N}$ vector models at large $N$, the quantum Lyapunov exponents are parametrically suppressed as $\lambda_L \sim T/N$ \cite{Swingle:1703.02545,Klug:1712.08813}.

Quantum entanglement between different degrees of freedom offers a complementary language for a quantitative description of quantum chaos. It is expected that for strongly interacting chaotic systems all degrees of freedom become highly entangled under quantum evolution \cite{Page:gr-qc/9305007,Susskind:0808.2096}, even if the initial state is a direct product $\ket{\Psi} = \ket{\Psi_A} \otimes \ket{\Psi_B}$ of states $\ket{\Psi_A}$ and $\ket{\Psi_B}$ of subsystems $A$ and $B$. The entanglement entropy is expected to exhibit a rapid growth at early times followed by saturation at late times, when the system has already ``scrambled'' the information contained in subsystem states $\ket{\Psi_A}$ and $\ket{\Psi_B}$ \cite{Susskind:0808.2096,Berenstein:1803.02396}. In the semiclassical approximation, the growth rate of entanglement entropy at early times is determined by the classical Lyapunov exponents \cite{Zurek:gr-qc/9402006,Berenstein:1503.04857}. However, beyond the semiclassical approximation the relation between Lyapunov exponents and growth of entanglement entropy could only be demonstrated for quadratic (or approximately quadratic) Hamiltonians \cite{Zurek:gr-qc/9402006,Berenstein:1503.04857,Bianchi:1709.00427} and for models with discrete time evolution \cite{Berenstein:1803.02396}.

In this paper we report on numerical studies of quantum corrections to the real-time dynamics of the thermal states of the BFSS model and its bosonic sector (bosonic matrix model), addressing in particular quantum corrections to Lyapunov exponents, the relation between Lyapunov exponents and entanglement entropy generation, and quasinormal ringing. We find that quantum corrections from the bosonic sector of the model tend to make the system less chaotic and less dissipative, whereas the contribution of Majorana fermions works in the opposite direction. The characteristic Lyapunov time, entanglement saturation time and decay time of quasinormal ringing become very long for the bosonic matrix model at sufficiently low temperatures, which roughly correspond to the confinement regime \cite{Hanada:1802.02985,Nishimura:0706.3517}. In contrast, for the full BFSS model with fermions these characteristic time scales remain finite even at the lowest energy accessible in our simulations. While at low temperatures at which the MSS bound is expected to be saturated our approximation is most likely too crude to capture the full dynamics of the model, our results suggest that quantum corrections from bosonic and fermionic sectors work in a way which is consistent with the MSS bound at lower temperatures, and which evades the naive violation of MSS bound by classical Lyapunov exponents $\lambda_L^0 \sim T^{1/4}$ \cite{Hanada:1512.00019} at sufficiently small $T$.

We further demonstrate that the characteristic saturation time for the entanglement entropy is in general shorter than the Lyapunov time $\tau_L \equiv \lambda_L^{-1}$ defined by the leading Lyapunov exponent $\lambda_L$. It appears to be governed by the classical, rather than quantum, leading Lyapunov exponent. The characteristic decay time of quasinormal ringing is found to be the shortest timescale of all.

In order to simulate the real-time dynamics of the BFSS model, we approximate the density matrix of the system by the most general Gaussian function with time-dependent parameters which obey self-consistent equations of motion. Such an approach, which we will refer to as the Gaussian state approximation, is closely related to the semiclassical approximation \cite{Heller:JChemPhys1975,Poulsen:AIP2017}, and is extensively used in the context of quantum chemistry  \cite{Heller:JChemPhys1975,Broeckhove:THEOCHEM199}. For interacting many-body systems which admit a second-quantized QFT description, such as e.g. the tight-binding description of electron gas in solids, time-dependent Gaussian state approximation for QFT is equivalent to the time-dependent Hartree-Fock approximation (see e.g. Chapter 12 of \cite{RingSchuckNuclearManyBody}) for the first-quantized many-body Hamiltonian.  For fermionic fields interacting with classical gauge fields, this approximation is equivalent to the classical-statistical field theory (CSFT) approximation which is by now a standard tool to study real-time dynamics of fermions interacting with highly occupied soft modes of gauge fields \cite{Berges:1403.4849}. An important property of the Gaussian state approximation is that it evolves pure states into pure states (see Appendix~\ref{apdx:symplectic_conservation} for the proof), which allows to study quantum entanglement in a consistent way.

As discussed in \cite{Heller:PhysicaD1992,Kaplan:nlin/0406054}, for classically chaotic systems the Gaussian state approximation, surprisingly, works even better than for systems which exhibit regular classical motion, and rather accurately describes the quantum evolution at the time scales of order of the classical Lyapunov time. Only some subtle late-time phenomena such as the wave-packet revival are not captured \cite{Hasegawa:1301.6423}. In \cite{Buividovich:17:5} we have also compared the Gaussian state approximation with the numerical solution of Schr\"{o}dinger equation for a simple classically chaotic Hamiltonian with two bosonic degrees of freedom \cite{Nicolai:NPB1989} which closely resembles the bosonic matrix model, and found a good agreement for evolution times $t \leq 2/\lambda_L^0$ less than approximately two Lyapunov times in both quantum and classical regimes. These observations suggest that the Gaussian state approximation should be at least qualitatively accurate for the description of real-time thermalization at time scales comparable with the classical Lyapunov time.

In the context of BFSS model, one of the limitations of the Gaussian state approximation is that the gauge symmetry constraints cannot be fully respected. As a consequence, our simulations correspond to the ungauged version of the BFSS or bosonic matrix models, where no gauge constraints are imposed on the state vectors. Fortunately, the differences between the gauged and ungauged models appear to be minor at least at low temperatures, as conjectured recently in \cite{Maldacena:1802.00428} and demonstrated numerically in \cite{Hanada:1802.02985}. Yet another argument in favor of accuracy of the Gaussian state approximation is that, as we will demonstrate, it reproduces the numerical results for the equation of state of the ungauged bosonic matrix model \cite{Hanada:1802.02985} within a few percent accuracy all the way from low to high temperatures.

We start our discussion in Section~\ref{sec:bfss_summary} by briefly reviewing the BFSS model and setting up the notations to be used in the rest of the paper. In Section~\ref{sec:gauss_approx} we explain the Gaussian state approximation for the real-time dynamics of the BFSS model. This approximation is rather general and can be easily extended to other models which admit Hamiltonian formulation. In Section~\ref{sec:thermal_initial} we discuss the initial state used in our simulations, which we require to resemble the thermal equilibrium state as close as possible. In Section~\ref{sec:num_res} we present our numerical results. In Subsection~\ref{subsec:lyapunov_mss} we demonstrate that quantum corrections make Lyapunov exponents smaller than in the classical system, thus being in agreement with the MSS bound (\ref{mss_bound_def}). We also clarify the relation of our results to out-of-time order correlators of the form (\ref{otoc_def}). In Subsection~\ref{subsec:entanglement_entropy} we study real-time evolution of entanglement entropy and discuss the relation between entanglement generation and Lyapunov exponents. In Subsection~\ref{subsec:qnfs} we consider quasinormal ringing and the temperature dependence of complex-valued quasinormal frequencies. In the concluding Section~\ref{sec:conclusions} we summarize our findings and outline some directions for further work. Technical details of our simulations are described in several Appendices.

\section{A brief review of the BFSS model}
\label{sec:bfss_summary}

In this paper we use the following representation of the Hamiltonian of the BFSS matrix model \cite{Susskind:hep-th/9610043}:
\begin{eqnarray}
\label{bfss_Hamiltonian}
 \hat{H}
 &=&
 \frac{\lambda}{2 N} \hP^a_i \hP^a_i
 +
 \frac{N}{4 \lambda} C_{abc} C_{ade} \hX^b_i \hX^c_j \hX^d_i \hX^e_j
\nonumber \\
& &
\qquad
+
 \frac{i}{2} \, C_{abc} \hpsi^a_{\alpha} \, \sigma_i^{\alpha\beta} \, \hX^b_{i} \hpsi^c_{\beta} .
\end{eqnarray}
In this expression and throughout the paper we use the following notations and conventions:
\begin{itemize}
 \item $\hX^a_i$ and $\hP^a_i$ are canonically conjugate bosonic coordinate and momentum operators with commutation relations $\lrs{\hX^a_i, \hP^b_j} = i \delta^{a b} \delta_{ij}$, which have dimensions of $\lr{Mass}$ and $\lr{Mass}^{-1}$, respectively. This is a natural convention because $X^a_i$ correspond in fact the components of the gauge field vector in $\lr{9+1}$-dimensional super-Yang-Mills theory.
 \item The indices $i, \, j, \, k, \, \ldots = 1 \ldots 9$ label the $d=9$ spatial coordinates.
 \item The indices $a, \, b, \, c, \, \ldots = 1 \ldots N^2-1$ label the elements of the $su\lr{N}$ Lie algebra - that is, the algebra of traceless Hermitian $N \times N$ matrices.
 \item $C_{a b c} = -i \tr\lr{T_a \lrs{T_b, T_c}}$ are the structure constants of the $su\lr{N}$ Lie algebra, with the generators $T_a$ normalized as $\tr\lr{T_a T_b} = \delta_{ab}$.
 \item $\lambda$ is the t'Hooft coupling constant which is kept fixed when taking the large-$N$ limit. The standard 't Hooft limit is realized by scaling  the energy to be of order $N^2$ as $N \rightarrow \infty$. $\lambda$ has a dimension of $\lr{Mass}^3$, and without loss of generality we can set it to unity by expressing all dimensionful quantities in units of $\lambda^{1/3}$: $\hX^a_i \rightarrow \lambda^{1/3} \hX^a_i$, $\hP^a_i \rightarrow \lambda^{-1/3} \hP^a_i$, $\hat{H} \rightarrow \lambda^{1/3} \hat{H}$.
 \item $\hpsi^a_{\alpha} = \lr{\hpsi^{a}_{\alpha}}^{\dag}$ are the dimensionless Majorana fermionic operators with anti-commutation relations $\lrc{\hpsi^a_{\alpha}, \hpsi^b_{\beta}} = \delta^{a b} \delta_{\alpha\beta}$. The indices $\alpha, \, \beta, \, \ldots$ run from 1 to 16. They correspond to the 16 elements of Weyl-Majorana spinors in $D=\lr{9+1}$ dimensions before dimensional reduction.
 \item $\sigma_i^{\alpha\beta}$, $i = 1 \ldots 9$ are the $d = 9$-dimensional analogues of the Pauli matrices, which are traceless, real, symmetric $16 \times 16$ matrices with anti-commutation relations $\sigma_i \sigma_j + \sigma_j \sigma_i = 2 \delta_{ij}$ (see Appendix~\ref{apdx:sigma_identities} for explicit construction and useful identities).
\end{itemize}
A nice summary of formulae for the BFSS model can be also found e.g. in \cite{Filev:1605.01611}.

Getting rid of explicit Lie algebra indices and treating $\hX_i$ and $\hpsi_{\alpha}$ as $N \times N$ Hermitian traceless matrices, we can also write the Hamiltonian (\ref{bfss_Hamiltonian}) as
\begin{eqnarray}
\label{bfss_Hamiltonian_matrix}
 \hat{H}
 &=&
 \frac{1}{2 N} \tr \hP_i^2
 -
 \frac{N}{4} \tr\lrs{\hX_i, \hX_j}^2
 \nonumber\\
 & &
 \qquad
 +
 \frac{\sigma_i^{\alpha\beta}}{2} \tr\lr{\hpsi_{\alpha} \lrs{\hX_i, \hpsi_{\beta} } },
\end{eqnarray}
where the commutators and traces are understood as operations on $N \times N$ matrices, rather than quantum-mechanical traces, and the t'Hooft coupling $\lambda$ is already set to unity.

The representation (\ref{bfss_Hamiltonian_matrix}) makes it obvious that the Hamiltonian (\ref{bfss_Hamiltonian}) is invariant under the simultaneous unitary similarity transformations of all the matrices $X_i$, $P_i$ and $\psi^a_{\alpha}$, which are generated by the operator
\begin{eqnarray}
\label{gauge_generators}
 \hat{J}^a = C_{abc} \hX^b_i \hP^c_i - \frac{i}{2} C_{abc} \hpsi^b_{\alpha} \hpsi^c_{\alpha}
\end{eqnarray}
acting as $\lrs{\hat{J}^a, \hat{O}^b} = i C_{abc} \hat{O}_c$ on any operator $\hat{O}^a$ which transforms under the adjoint representation of $SU\lr{N}$, e.g. $\hX^a_i$, $\hP^a_i$, $\hpsi^a_{\alpha}$. This symmetry is a remnant of the gauge symmetry of the $\lr{9 + 1}$ dimensional super-Yang-Mills theory, from which the BFSS Hamiltonian (\ref{bfss_Hamiltonian}) can be obtained by dimensional reduction. Correspondingly, the physical Hilbert space is defined by imposing the constraint $\hat{J}^a \ket{\Psi} = 0$ on physical states.

On the space of physical states defined by $\hat{J}^a \ket{\Psi} = 0$ the BFSS Hamiltonian (\ref{bfss_Hamiltonian}) also commutes with $\mathcal{N}=16$ supersymmetry generators
\begin{eqnarray}
\label{bfss_susy_charge}
 \hat{Q}^{\alpha}
 =
 \hP^a_i \, \sigma_i^{\alpha\beta} \, \hpsi^a_{\beta}
 -
 \frac{N}{4} C_{abc} \hX^b_i \hX^c_j \, \sigma_{ij}^{\alpha\beta} \, \hpsi^a_{\beta} ,
\end{eqnarray}
where $\sigma_{ij} \equiv \sigma_i \sigma_j - \sigma_j \sigma_i$.

\section{Gaussian state approximation for the real-time dynamics of the BFSS model}
\label{sec:gauss_approx}

In this section, we explain how the Gaussian state approximation is obtained by truncating the full equations of motion (Heisenberg equations) ${\partial_t \hat{O} = i \lrs{\hat{H}, \hat{O}}}$ for the canonical coordinate operators $\hX^a_i$, $\hP^a_i$ and $\hpsi^a_{\alpha}$:
\begin{subequations}
\label{heisenberg_eqs}
\begin{eqnarray}
 \label{heisenberg_eqs_dtX}
 \partial_t \hX^a_i
 &=&
 \frac{1}{N} \, \hP^a_i, \quad
 \\ {} \nonumber \\
 \label{heisenberg_eqs_dtP}
 \partial_t \hP^a_i
 &=&
 - N \, C_{abc} C_{cde} \hX^b_j \hX^d_i \hX^e_j
 -
 \frac{i}{2} C_{b a c} \sigma_i^{\alpha\beta} \hpsi^b_{\alpha} \hpsi^c_{\beta},  \quad
 \\ {} \nonumber \\
 \label{heisenberg_eqs_ferm}
 \partial_t \hpsi^a_{\alpha}
 &=&
 C_{a b c} \hX^b_i \sigma_i^{\alpha\beta} \hpsi^c_{\beta} . \quad
\end{eqnarray}
\end{subequations}

Averaging these equations of motion over some density matrix, we can express the time derivatives of the expectation values $\vev{\hX^a_i}$ and $\vev{ \hP^a_i }$ in terms of equal-time correlators of up to three operators $\hX$, $\hP$ and/or $\hpsi$. The equations of motion for these correlators would include correlators with even larger number of operators, and we would obtain an infinite hierarchy of equations similar to the Schwinger-Dyson equations which cannot be treated neither numerically nor analytically without further approximations.

In order to obtain a treatable approximation to the full Heisenberg equations (\ref{heisenberg_eqs}) which involves only a finite number of variables, let us restrict the time-dependent density matrix $\bra{X, \psi} \hat{\rho} \ket{X', \psi'}$ to be the most general Gaussian functional of $X$, $X'$, $\psi$ and $\psi'$ with time-dependent parameters \cite{Heller:JChemPhys1975,Broeckhove:THEOCHEM199} (where $\ket{X, \psi}$ are the eigenstates of the bosonic and fermionic operators $\hX^a_i$ and $\hpsi^a_{\alpha}$). In other words, the matrix elements $\bra{X, \psi} \hat{\rho} \ket{X', \psi'}$ can be represented as $\mathcal{N} \, \expa{-F\lr{X, \psi, X', \psi'}}$, where $F\lr{X, \psi, X', \psi'}$ is the most general quadratic polynomial of $X$, $X'$, $\psi$ and $\psi'$ with a time-dependent normalization factor $\mathcal{N}$. Such Gaussian density matrices can be unambiguously parameterized in terms of one- and two-point correlators of canonical variables due to Wick's theorem.
Therefore, by using the Gaussian density matrix, all the equal-time correlators of the canonical variables $\hX^a_i$, $\hP^a_i$ and $\hpsi^a_{\alpha}$ are expressed in terms of one-point and two-point correlators. Let us introduce the following concise notation:
\begin{eqnarray}
\label{correlators_def}
 X^a_i
 &\equiv&
  \vev{\hX^a_i} \equiv \tr\lr{ \hat{\rho} \, \hX^a_i },
 \nonumber \\
 P^a_i
 &\equiv&
  \vev{\hP^a_i} \equiv \tr\lr{ \hat{\rho} \, \hP^a_i },
 \nonumber \\ {} \nonumber \\
 \cev{\hX^a_i \hX^b_j}
 &\equiv&
 \vev{\hX^a_i \hX^b_j} - \vev{\hX^a_i} \vev{\hX^b_j}
 \nonumber \\
 &\equiv&
 \tr\lr{ \hat{\rho} \, \hX^a_i \hX^b_j }
 -
 \tr\lr{ \hat{\rho} \, \hX^a_i}
 \tr\lr{ \hat{\rho} \, \hX^b_j} ,
 \nonumber \\ {} \nonumber \\
 \cev{\hP^a_i \hP^b_j}
 &\equiv&
 \vev{\hP^a_i \hP^b_j} - \vev{\hP^a_i} \vev{\hP^b_j}
\nonumber \\
&\equiv&
 \tr\lr{ \hat{\rho} \, \hP^a_i \hP^b_j }
 -
 \tr\lr{ \hat{\rho} \, \hP^a_i}
 \tr\lr{ \hat{\rho} \, \hP^b_j} ,
 \nonumber \\ {} \nonumber \\
 \cev{\hX^a_i \hP^b_j}
 &\equiv&
 \vev{\hX^a_i \hP^b_j} - \vev{\hX^a_i} \vev{\hP^b_j}
  \nonumber \\
  &\equiv&
 \frac{1}{2} \, \tr\lr{ \hat{\rho} \, \lr{\hX^a_i \hP^b_j + \hP^b_j \hX^a_i} }
\nonumber \\
 & &-
 \tr\lr{ \hat{\rho} \, \hX^a_i}
 \tr\lr{ \hat{\rho} \, \hP^b_j} .
\end{eqnarray}
Symmetrization of the product of $\hX^a_i$ and $\hP^b_j$ operators in the last definition ensures the real-valuedness of the equal-time correlator $\cev{X^a_i P^b_j}$, and also allows us to work with Wigner functions in a more straightforward way (see below). While it is possible to introduce the mixed bosonic-fermionic correlators of the form $\vev{\hpsi \hX}$, $\vev{\hpsi \hP}$ and fermionic one-point functions $\vev{\hpsi}$ as well, one can straightforwardly demonstrate that if they vanish in the initial state, they remain zero during all the subsequent evolution. Since states with nonzero expectation values $\vev{\hpsi \hX}$, $\vev{\hpsi \hP}$ and $\vev{\hpsi}$ are rather exotic excited states, we restrict our analysis to initial states where only the correlators (\ref{correlators_def}) are nonzero. We note that these correlators can also be put in one-to-one correspondence with the Green functions $G^{++} \sim \vev{X\lr{t^+} X\lr{t^+}}$, $G^{+-} \sim \vev{X\lr{t^+} X\lr{t^-}}$ and $G^{--} \sim \vev{X\lr{t^-} X\lr{t^-}}$ on the Keldysh contour parameterized by time variables $t^+$ and $t^-$ on the forward and backward branch, respectively. Throughout the paper we will often refer to $X^a_i$ and $P^a_i$ as the classical coordinates and momenta. This interpretation is justified when $\exp(-F\lr{X, \psi, X', \psi'})$ is sufficiently localized.

Averaging equations (\ref{heisenberg_eqs}) over the Gaussian density matrix characterized by the correlators (\ref{correlators_def}) and applying Wick's theorem, we obtain the following equations for the time evolution of $X^a_i$ and $P^a_i$:
\begin{subequations}
\label{gs_onepoint_eqs}
\begin{eqnarray}
 \label{gs_onepoint_eq_X}
 \partial_t X^a_i
 =
  \frac{1}{N} P^a_i,
\end{eqnarray}
\begin{eqnarray}
 \label{gs_onepoint_eq_P}
 \partial_t P^a_i
 &=&
  - N C_{abc} C_{cde} X^b_j X^d_i X^e_j
  \nonumber \\
  & &
 -
 N C_{abc} C_{cde} X^b_j \cev{\hX^d_i \hX^e_j}
  \nonumber \\
  & &
 -
 N C_{abc} C_{cde} \cev{\hX^b_j \hX^e_j} X^d_i
  \nonumber \\
  & &
 -
 N C_{abc} C_{cde} \cev{\hX^b_j \hX^d_i} X^e_j
  \nonumber \\
  & &
 -
 \frac{i}{2} C_{b a c} \sigma^i_{\alpha\beta} \cev{\hpsi^b_{\alpha} \hpsi^c_{\beta}} .
\end{eqnarray}
\end{subequations}
To make our approximation self-consistent, we also need to describe the time evolution of the two-point correlators which enter equations (\ref{gs_onepoint_eqs}). To this end let us write down the Heisenberg equations governing the time evolution of the composite operators $\hX^a_i \hX^b_j$, $\frac{1}{2}\lr{\hX^a_i \hP^b_j + \hP^b_j \hX^a_i}$, $\hP^a_i \hP^b_j$ and $\hpsi^a_{\alpha} \hpsi^b_{\beta}$:
\begin{subequations}
\label{heisenberg_eqs_twopoint}
\begin{eqnarray}
 \label{heisenberg_eqs_twopoint_XX}
 \partial_t \lr{\hX^a_i \hX^b_j}
 = \lr{\hP^a_i \hX^b_j + \hX^a_i \hP^b_j}/N ,
 \end{eqnarray}
 \begin{eqnarray}
 \label{heisenberg_eqs_twopoint_XP}
 \partial_t \lr{\hP^a_i \hX^f_k + \hX^f_k \hP^a_i}/2
 = \nonumber\\ = 
 \hP^a_i \hP^f_k / N
 -
 N C_{abc} C_{cde} \hX^b_j \hX^d_i \hX^e_j \hX^f_k
 - \nonumber \\ -
 \frac{i}{2}
 C_{bac} \sigma_i^{\alpha\beta} \hpsi^b_{\alpha} \hpsi^c_{\beta} \hX^f_k ,
\end{eqnarray}
\begin{eqnarray}
\label{heisenberg_eqs_twopoint_PP}
 \partial_t \lr{\hP^a_i \hP^f_k}
 =
 - N C_{abc} C_{cde} \hX^b_j \hX^d_i \hX^e_j \hP^f_k
 - \nonumber \\ - 
 \frac{i}{2} C_{bac} \sigma_i^{\alpha\beta} \hpsi^b_{\alpha} \hpsi^c_{\beta} \hP^f_k
 + \nonumber\\ +
 \lr{ \lrc{a,i} \leftrightarrow \lrc{f, k} } ,
\end{eqnarray}
\begin{eqnarray}
\label{heisenberg_eqs_twopoint_psipsi}
 \partial_t \lr{\hpsi^a_{\alpha} \hpsi^d_{\gamma}}
=
 C_{abc} \hX^b_i \sigma_i^{\alpha\beta} \hpsi^c_{\beta} \hpsi^d_{\gamma}
 + \nonumber\\ +
 C_{dbc} \hX^b_i \sigma_i^{\gamma\beta} \hpsi^a_{\alpha} \hpsi^c_{\beta} .
\end{eqnarray}
\end{subequations}
We can again average these equations over our Gaussian density matrix and apply Wick's theorem. This is straightforward for all equations except (\ref{heisenberg_eqs_twopoint_PP}), where one has to express the expectation values of the form $\vev{\hX^b_j \hX^d_i \hX^e_j \hP^f_k}$ in terms of two-point functions (\ref{correlators_def}).
Since $\hX$ and $\hP$ do not commute, one cannot treat them as ordinary commuting numbers, and the application of Wick's theorem is not straightforward. Indeed, when averaging all other equations we have implicitly used a representation of the density matrix $\rho$ in terms of the eigenstates of either $\hX$ or $\hP$ operators, which is obviously a Gaussian functional in both cases. Such a representation cannot be used for correlators which contain both $\hX$ and $\hP$ operators. A simple solution to this problem is to use the Wigner transform
\begin{eqnarray}
\label{dm_wigner_transform}
 \rho\lr{X, P} = \int dY e^{-i P \cdot Y} \bra{X + Y/2} \hat{\rho} \ket{X - Y/2}
\end{eqnarray}
of the density matrix $\hat{\rho}$. If $\bra{X} \hat{\rho} \ket{X'}$ is a Gaussian functional of $X$ and $X'$, the Wigner transform (\ref{dm_wigner_transform}) is also a Gaussian functional of $X$ and $P$. Using the definition (\ref{dm_wigner_transform}), one can show that the ``classical'' phase space integrals of the form $\int dX \, dP \, \rho\lr{X, P} \, O\lr{X} \, P^a_i$ are related to vacuum expectation value of the symmetrized operator product
\begin{eqnarray}
\label{Wigner_vev}
 \int dX \, dP \, \rho\lr{X, P} \, O\lr{X} \, P^a_i
 = \nonumber\\ = 
 \frac{1}{2}\tr\lr{\hat{\rho} \, \lr{O(\hX) \, \hP^a_i + \hP^a_i \, O(\hX)}} ,
\end{eqnarray}
where $O(\hX)$ can be any operator which commutes with all operators $\hX^a_i$. Since the right-hand side of the equation (\ref{Wigner_vev}) is a Gaussian integral over ordinary commuting variables, we can apply Wick's theorem to the symmetrized operator products like the ones on the left-hand side of (\ref{Wigner_vev}).

Since we have assumed that the only nonzero correlator with fermions is $\vev{\hpsi^a_{\alpha} \hpsi^b_{\beta}}$ and $\vev{\hpsi} = 0$, $\vev{\hpsi \hX} = 0$, $\vev{\hpsi \hP} = 0$, fermionic terms in our Gaussian density matrix completely decouple from the bosonic ones, and can be safely disregarded in the above considerations. In fact, we don't even need Wick's theorem for fermions, since correlators with more than two fermionic operators never appear in our equations of motion.

Symmetrizing the operator products in the expectation values $\vev{\hX^b_j \hX^d_i \hX^e_j \hP^f_k}$ in (\ref{heisenberg_eqs_twopoint_PP}) and convoluting all equations (\ref{heisenberg_eqs_twopoint}) with a Gaussian Wigner transform of the density matrix $\hat{\rho}$, we obtain the following equations for the time evolution of the two-point correlators in (\ref{correlators_def}):
\begin{subequations}
\label{gs_twopoint_eqs}
\begin{eqnarray}
 \label{gs_twopoint_eqs_XX}
 \partial_t \cev{\hX^a_i \hX^b_j} = \frac{\cev{\hX^a_i \hP^b_j} + \cev{\hX^b_j \hP^a_i}}{N},
\end{eqnarray}
\begin{eqnarray}
 \label{gs_twopoint_eqs_XP}
 \partial_t \cev{\hX^f_k \hP^a_i }
 =
 \cev{\hP^a_i \hP^f_k}/N
 - \nonumber \\ -
 N C_{abc} C_{cde} \vev{\hX^d_i \hX^e_j} \cev{\hX^b_j \hX^f_k}
 - \nonumber \\ -
 N C_{abc} C_{cde} \vev{\hX^b_j \hX^e_j} \cev{\hX^d_i \hX^f_k}
 - \nonumber \\ -
 N C_{abc} C_{cde} \vev{\hX^b_j \hX^d_i} \cev{\hX^e_j \hX^f_k} ,
\end{eqnarray}
\begin{eqnarray}
 \label{gs_twopoint_eqs_PP}
 \partial_t \cev{\hP^a_i \hP^f_k}
 = \nonumber \\ =  
 - N C_{abc} C_{cde} \vev{\hX^d_i \hX^e_j} \cev{\hX^b_j \hP^f_k}
 - \nonumber \\ -
 N C_{abc} C_{cde} \vev{\hX^b_j \hX^e_j} \cev{\hX^d_i \hP^f_k}
- \nonumber \\ -
 N C_{abc} C_{cde} \vev{\hX^b_j \hX^d_i} \cev{\hX^e_j \hP^f_k}
+ \nonumber \\ +
 \lr{ \lrc{a,i} \leftrightarrow \lrc{f, k} } ,
 \end{eqnarray}
\begin{eqnarray}
 \label{gs_twopoint_eqs_psipsi}
 \partial_t \cev{\hpsi^a_{\alpha} \hpsi^d_{\gamma}}
  = 
 C_{abc} X^b_i \sigma_i^{\alpha\beta} \cev{ \hpsi^c_{\beta} \hpsi^d_{\gamma} }
+ \nonumber \\ +
 C_{dbc} X^b_i \sigma_i^{\gamma\beta} \cev{ \hpsi^a_{\alpha} \hpsi^c_{\beta} } .
\end{eqnarray}
\end{subequations}
Note that equations of motion for the bosonic two-point correlators do not contain fermionic correlators. Fermions only affect the dynamics due to the coupling to the expectation values $X^a_i \equiv \vev{\hX^a_i}$, which enter the equations (\ref{gs_twopoint_eqs_XX}), (\ref{gs_twopoint_eqs_XP}) and (\ref{gs_twopoint_eqs_PP}) via the disconnected correlators $\vev{\hX^a_i \hX^b_j} \equiv \cev{\hX^a_i \hX^b_j} + \vev{\hX^a_i} \vev{\hX^b_j}$.

Equations (\ref{gs_onepoint_eqs}) and (\ref{gs_twopoint_eqs}) provide a full and consistent system of equations for the time evolution of the correlators (\ref{correlators_def}). In particular, one can show that these equations conserve the expectation values of the Hamiltonian (\ref{bfss_Hamiltonian}) and the angular momentum (\ref{angular_momentum}), provided these are also expressed in terms of the correlators (\ref{correlators_def}) using Wick's theorem. Explicit expressions for these conserved quantities are given in Appendix~\ref{apdx:csft_symmetries}. On the other hand, supersymmetry generators (\ref{bfss_susy_charge}) are not conserved, see Appendix~\ref{apdx:csft_symmetries} for a detailed discussion.

The conservation of the generators of the gauge transformations (\ref{gauge_generators}) is important for what follows and requires a special discussion. One can show that, similarly to the energy and the angular momentum, equations (\ref{gs_onepoint_eqs}) and (\ref{gs_twopoint_eqs}) conserve the expectation values $\vev{\hat{J}^a} = \bra{\Psi} \hat{J}^a \ket{\Psi}$ of the gauge constraint, which we require to vanish in the initial state of our system. The gauge constraint $\hat{J}^a \ket{\Psi} = 0$ in the full quantum treatment is, however, much stronger, and is equivalent to the vanishing of $\bra{\Psi'} \hat{J}^a \ket{\Psi}$ for an arbitrary state vector $\ket{\Psi'}$. It is straightforward to check that there is no normalizable Gaussian wave function $\ket{\Psi}$ which satisfies the equation $\hat{J}^a \ket{\Psi} = 0$. The Gaussian state approximation is thus only able to describe the ungauged versions of the BFSS model and the bosonic matrix model. Since the BFSS model is only supersymmetric on the space of gauge-invariant states, we also conclude that supersymmetry cannot be preserved within the Gaussian state approximation (see Appendix~\ref{apdx:csft_symmetries} for a more detailed discussion).

Fortunately, as discussed recently in \cite{Maldacena:1802.00428,Hanada:1802.02985}, the physics of the bosonic matrix model and the BFSS model does not strongly depend on gauging.
More precisely, gauged and ungauged theories are expected to be the same up to the $e^{-C/T}$ correction at low temperature, where $C$ is an order one constant. Their behavior in the high-temperature region is also qualitatively the same; in particular, the real-time aspects in the gauge singlet sector are exactly the same in the high temperature limit, if the energies are taken to be the same. The description of the ungauged bosonic matrix model within the Gaussian state approximation appears to be rather good, as suggested by the comparison of the thermodynamic equation of state with numerical data of \cite{Hanada:1802.02985} in Section~\ref{sec:thermal_initial}.

Another important property of equations (\ref{gs_onepoint_eqs}) and (\ref{gs_twopoint_eqs}) is that they evolve pure states into pure states, and, more generally, conserve the von Neumann entropy of the density matrix $\hat{\rho}$ (see Appendix~\ref{apdx:symplectic_conservation}). This allows to study quantum entanglement between different degrees of freedom in a meaningful way, see Subsection~\ref{subsec:entanglement_entropy}. Still, one has to keep in mind that the Gaussian state approximation does not describe a unitary evolution. In particular, scalar products between different Gaussian states and expectation values like $\vev{\hat{H}^2}$ should be conserved for unitary evolution described by the operator $e^{i \hat{H} t}$, but are not conserved within the Gaussian state approximation.
This is because the energy eigenstates are not necessarily Gaussian.

Similarly to the full Schr\"{o}dinger equation ${\partial_t \ket{\Psi} = i \hat{H} \ket{\Psi}}$, which can be obtained by extremizing the ``quantum'' action $S_q = \int dt \bra{\Psi} \lr{\partial_t - i \hat{H}} \ket{\Psi}$ over all possible time histories of a unit state vector $\ket{\Psi}$, equations (\ref{gs_onepoint_eqs}) and (\ref{gs_twopoint_eqs}) can be obtained by restricting this extremization to the space of all possible time-dependent Gaussian states \cite{Broeckhove:THEOCHEM199}. One can also interpret (\ref{gs_onepoint_eqs}) and (\ref{gs_twopoint_eqs}) as classical equations of motion which follow from a certain extension of the classical Hamiltonian \cite{Heller:JChemPhys1975,Broeckhove:THEOCHEM199}. This property allows one to identify a symplectic structure of these equations (see Appendix~\ref{apdx:symplectic_conservation}) and devise stable leapfrog-type numerical integrators.

In contrast to the full Schr\"{o}dinger equation for the Hamiltonian (\ref{bfss_Hamiltonian}), equations (\ref{gs_onepoint_eqs}) and (\ref{gs_twopoint_eqs}) contain a finite number of variables which scales only polynomially with the number of degrees of freedom, which allows for an efficient numerical solution even for large physical systems. In particular, this mild scaling is a motivation for using the Gaussian state approximation to study quantum real-time dynamics in quantum chemistry \cite{Heller:JChemPhys1975,Broeckhove:THEOCHEM199}. In our case, the most computationally intensive part of the simulations is the solution of equations (\ref{gs_twopoint_eqs_XP}) and (\ref{gs_twopoint_eqs_PP}). CPU time usage is dominated by the calculation of the terms $C_{abc} C_{cde} \vev{\hX^b_j \hX^d_i} \cev{\hX^e_j \hX^f_k}$ and $C_{abc} C_{cde} \vev{\hX^b_j \hX^d_i} \cev{\hX^e_j \hP^f_k}$ on the right-hand side of (\ref{gs_twopoint_eqs_XP}) and (\ref{gs_twopoint_eqs_PP}). The structure of Wick contractions in these terms neither allows to use the functions which calculate a single commutator, nor to save time by contracting some of the spatial indices prior to contracting the matrix indices. Naively, index contractions in these terms require $O\lr{d^3 N^6}$ floating-point operations. By explicitly taking into account the structure of $C_{abc}$ tensors we have achieved an $O\lr{d^3 N^5}$ scaling, which is still significantly more dramatic than the $O\lr{d^2 N^3}$ scaling for the simulations of the classical dynamics, and thus significantly limits the range of accessible $N$ values.

In this work we consider three different approximations to the full real-time dynamics of the BFSS model (\ref{bfss_Hamiltonian}), all of which can be obtained from equations (\ref{gs_onepoint_eqs}) and (\ref{gs_twopoint_eqs}):
\begin{enumerate}[label={\arabic*)}]
 \item \textbf{Classical dynamics} of the Hamiltonian (\ref{bfss_Hamiltonian}). The corresponding equations of motion are obtained from equations (\ref{gs_onepoint_eqs}) by setting all two-point correlators to zero. Since there is no classical limit for fermionic dynamics, fermions are completely neglected in this approximation. The classical approximation becomes quantitatively exact both for the bosonic matrix model and the full BFSS model at asymptotically high energies/temperatures.
 \item Real-time dynamics of the ungauged \textbf{bosonic matrix model} in the Gaussian state approximation, which corresponds to the Hamiltonian (\ref{bfss_Hamiltonian}) and equations (\ref{gs_onepoint_eqs}) and (\ref{gs_twopoint_eqs}) without fermionic terms. In contrast to the full BFSS model, at low temperatures the bosonic matrix model is expected to be in the confinement regime \cite{Hanada:0707.4454} with finite ground-state energy. While in the ungauged bosonic matrix model there is no strict notion of confinement and the high- and low-temperature regimes appear to be smoothly connected, at sufficiently low temperatures physical observables in the gauged and in the ungauged models become exponentially close \cite{Maldacena:1802.00428,Hanada:1802.02985}.
 \item Real-time dynamics of the full ungauged \textbf{BFSS model} in the Gaussian state approximation.
\end{enumerate}

\section{Thermal initial conditions and equation of state}
\label{sec:thermal_initial}

Equations (\ref{gs_onepoint_eqs}) and (\ref{gs_twopoint_eqs}) which approximately describe the real-time dynamics of the BFSS model should still be supplemented with suitable initial conditions.
In this paper, we are mostly interested in the real-time responses of thermal and nearly thermal states. Hence we take the initial conditions to reproduce the properties of thermal equilibrium states of the ungauged bosonic matrix model and the ungauged BFSS model as close as possible.
In this Section we explicitly construct such initial conditions within the Gaussian state approximation.

\subsection{Bosonic matrix model}
\label{subsec:eos_bosonic}

Within the Gaussian state approximation the thermal density matrix by definition should also be Gaussian  (i.e. correspond to a Gaussian Wigner function). If the system is in contact with a thermostat which does not perform work (e.~g.~ collisions with a hard wall), upon thermalization the von Neumann entropy of a state should reach its maximal possible value for a given energy. Based on this very general physical principle, we will approximate the thermal equilibrium states by those Gaussian density matrices which have the largest possible von Neumann entropy at a given energy.

To this end we need to know the von Neumann entropy of an arbitrary Gaussian density matrix, which can be expressed in terms of the correlators (\ref{correlators_def}). This relation has been addressed in detail in \cite{Sorkin:PRD1986}, and more recently in \cite{Sorkin:1205.2953,Sorkin:1311.7146,Berenstein:1503.04857,Berges:1712.09362}, where it was demonstrated that the von Neumann entropy of a Gaussian density matrix can be expressed in terms of the so-called symplectic eigenvalues of the block matrix
\begin{eqnarray}
\label{correlator_block_matrix}
 \Delta = \left(
   \begin{array}{cc}
     \cev{\hX^a_i \hX^b_j} & \cev{\hX^a_i \hP^b_j} \\
     \cev{\hX^b_j \hP^a_i} & \cev{\hP^a_i \hP^b_j} \\
   \end{array}
 \right)
\end{eqnarray}
of the size $2 N_{tot} \times 2 N_{tot}$, where $N_{tot} = d \lr{N^2 - 1}$ is the total number of bosonic degrees of freedom in our system.

Symplectic eigenvalues of the matrix (\ref{correlator_block_matrix}) are related to the eigenvalues of the matrix $\Delta \Omega$, where $\Omega$ is the symplectic form for the canonical coordinates $X^a_i$, $P^b_j$:
\begin{eqnarray}
\label{symplectic_form_def}
 \Omega = \left(
   \begin{array}{cc}
     0 & \delta^{ab} \delta_{ij} \\
     -\delta^{ab} \delta_{ij} & 0 \\
   \end{array}
 \right) .
\end{eqnarray}
For any positive-definite correlator matrix of the form (\ref{correlator_block_matrix}) the eigenvalues $\lambda$ of the matrix $\Delta \Omega$ come in complex conjugate pairs of the form $\lambda_{2 k - 1} = + i f_k$, $\lambda_{2 k} = - i f_k$. The real and positive numbers $f_k$, $k = 1 \ldots N_{tot}$ are called symplectic eigenvalues of $\Delta$. Quantum uncertainty relations imply that $f_k \geq 1/2$. A necessary and sufficient condition for the correlator matrix (\ref{correlator_block_matrix}) to describe a pure Gaussian state is that $f_k = 1/2$ for all $k$. It is easy to check that for a single bosonic coordinate $\hat{x}$ this identity implies $\cev{\hat{x}^2} \cev{\hat{p}^2} - \cev{\hat{x} \hat{p}}^2 = 1/4$. In other words, the Heisenberg uncertainty relation should be saturated. In Appendix~\ref{apdx:symplectic_conservation} we demonstrate that equations (\ref{gs_twopoint_eqs}) conserve symplectic eigenvalues $f_k$ and thus map pure states to pure states.

The von Neumann entropy $S = -\tr\lr{\hat{\rho} \ln \hat{\rho}}$ of a Gaussian state characterized by the correlator matrix (\ref{correlator_block_matrix}) can be expressed in terms of symplectic eigenvalues $f_k$ as \cite{Sorkin:PRD1986,Sorkin:1205.2953,Sorkin:1311.7146,Berges:1712.09362}
\begin{eqnarray}
\label{gs_von_Neumann}
 S
 &=&
 \sum\limits_k
 \lr{f_k + \frac{1}{2}} \ln\lr{f_k + \frac{1}{2}}
 \nonumber \\
 & &
 -
 \sum\limits_k
 \lr{f_k - \frac{1}{2}} \ln\lr{f_k - \frac{1}{2}} .
\end{eqnarray}
As it should be, the von Neumann entropy is equal to zero for pure states, and positive for mixed states. In the classical limit, when the $f_k$ are large, the von Neumann entropy approaches the classical entropy and can be expanded as
\begin{eqnarray}
\label{quantum_entropy_expansion}
 S = \sum\limits_{k} \ln\lr{f_k} + 1 + O\lr{f_k^{-2}} .
\end{eqnarray}

Thermal equilibrium states should be invariant under spatial and internal $SU\lr{N}$ rotations, as well as under discrete time-reversal and parity transformations. These symmetries imply $X^a_i = 0$, $P^a_i = 0$ and the following form of the correlators (\ref{correlators_def}):
\begin{eqnarray}
\label{bfss_exact_initial}
\cev{\hpsi^a_{\alpha} \hpsi^b_{\beta}}
&\sim&
\delta^{ab} \delta_{\alpha\beta},
 \nonumber \\
 \cev{\hX^a_i \hP^b_j}
 &=&
  0,
 \nonumber \\
 \cev{\hX^a_i \hX^b_j}
 &=&
  \xx \, \delta_{ij} \delta^{ab},
 \nonumber \\
 \cev{\hP^a_i \hP^b_j}
 &=&
  \pp \, \delta_{ij} \delta^{ab} .
\end{eqnarray}

The von Neumann entropy (\ref{gs_von_Neumann}) for the Gaussian density matrix characterized by such correlators is given by
\begin{eqnarray}
\label{equilibrium_von_Neumann}
 S
 &=&
 d \lr{N^2 - 1} \left(
  \lr{f + \frac{1}{2}}\ln\lr{f + \frac{1}{2}}
  \right. \nonumber \\
  & &
  \qquad\qquad\qquad\left. -
 \lr{f - \frac{1}{2}}\ln\lr{f - \frac{1}{2}}
 \right) ,
\end{eqnarray}
where $f = \sqrt{\xx \xp}$ is the $d \lr{N^2 - 1}$-fold degenerate symplectic eigenvalue of the block matrix (\ref{correlator_block_matrix}) constructed from correlators (\ref{bfss_exact_initial}).

Substituting the correlators (\ref{bfss_exact_initial}) into the expression (\ref{gs_conserved_energy}), we also obtain the corresponding energy
\begin{eqnarray}
\label{equilibrium_energy_vev}
 E = d \lr{N^2 - 1} \lr{\frac{\pp}{2 N} + \frac{N^2 \xx^2 \lr{d-1}}{2}} .
\end{eqnarray}

In order to maximize the entropy (\ref{equilibrium_von_Neumann}) at fixed energy $E$, we use equation (\ref{equilibrium_energy_vev}) to express $\pp$ in terms of $\xx$ and $E$, which yields
\begin{eqnarray}
\label{equilibrium_f}
 f^2 = \xx \pp = \frac{2 E \xx}{d} \frac{N}{N^2 - 1} - N^3 \xx^3 \lr{d - 1} .
\end{eqnarray}
Since the entropy (\ref{equilibrium_von_Neumann}) is a monotonically increasing function of $f$, it has a maximum with respect to $\xx$ when the equation $\partial f^2/\partial \xx = 0$ is satisfied, which yields
\begin{eqnarray}
\label{equilibrium_vals}
 E
 &=&
  \frac{3 d \lr{d-1}}{2} N^2 \lr{N^2 - 1} \xx^2 ,
 \nonumber \\
 f^2
 &=&
  2 N^3 \xx^3 \lr{d-1} .
\end{eqnarray}
It is now convenient to express physical observables in terms of the equilibrium value of $f$ given by (\ref{equilibrium_vals}):
\begin{eqnarray}
 \xx
 &=&
  N^{-1} f^{2/3} \lr{2 d - 2}^{-1/3} ,
 \nonumber \\
 \pp
 &=&
  N f^{4/3} \lr{2 d - 2}^{1/3} ,
 \nonumber \\
 E/N^2
 &=&
  \frac{3}{4} d \lr{2 d- 2}^{1/3} f^{4/3} \lr{1 - \frac{1}{N^2}} ,
 \nonumber \\
\left\langle\frac{1}{N} \tr\lr{\hX_i^2}\right\rangle
 &=&
 d \, \frac{N^2 - 1}{N} \, \xx = 
\nonumber \\
&=&
 d \lr{2 d - 2}^{-1/3} \, f^{2/3} \lr{1 - \frac{1}{N^2}} .
 \nonumber\\
 \label{equilibrium_vals_vs_f}
\end{eqnarray}
One can immediately check that the correlators (\ref{bfss_exact_initial}) with $\xx$ and $\pp$ given by (\ref{equilibrium_vals_vs_f}) provide a time-independent solution of equations (\ref{gs_onepoint_eqs}) and (\ref{gs_twopoint_eqs}), as it should be for thermal equilibrium states.

Now the only missing ingredient in our analysis of the equation of state is the temperature, which can be introduced using the standard thermodynamic relation $T^{-1} = \frac{\partial S}{\partial E}$. Expressing this derivative in terms of partial derivatives with respect to $f$, we obtain
\begin{eqnarray}
\label{temperature_def}
 T
 =
 \frac{\partial E}{\partial f}\cdot \left( \frac{\partial S}{\partial f}\right)^{-1}
 =
 \frac{\lr{2 d - 2}^{1/3} f^{1/3}}{\ln\lr{\frac{f + 1/2}{f - 1/2}}} .
\end{eqnarray}
This equation, together with (\ref{equilibrium_vals_vs_f}), provides a parametric form of the equation of state, and allows to express energy and other physical quantities such as $\frac{1}{N}\vev{ \tr\hX_i^2}$ in terms of the temperature $T$. Taking the high-temperature limit which is equivalent to the large-$f$ limit, we reduce equation (\ref{temperature_def}) to the form
\begin{eqnarray}
\label{temperature_def_highT}
 T = \lr{2 d - 2}^{1/3} f^{4/3} .
\end{eqnarray}
This immediately leads to the high-temperature asymptotics of the equation of state
\begin{eqnarray}
\label{classic_ungauged_eos}
 E = \frac{3}{4} \, d \, \lr{N^2 - 1} \, T ,
\end{eqnarray}
which is exactly the classical equation of state for the ungauged bosonic matrix model \cite{Maldacena:1802.00428,Hanada:1802.02985}. It can also be obtained by replacing the quantum von Neumann entropy (\ref{equilibrium_von_Neumann}) with the corresponding classical expression $S = d \lr{N^2 - 1} \ln\lr{f} + \const$. The relations (\ref{equilibrium_vals_vs_f}) and (\ref{temperature_def_highT}) also allow to express the coordinate dispersion at asymptotically high temperatures as
\begin{eqnarray}
\label{classic_relations}
 \xx
 &=&
 N^{-1} T^{1/2} \lr{2 d - 2}^{-1/2},
 \nonumber \\
 \frac{1}{N}\vev{ \tr X_i^2}
 &=&
 d \cdot\frac{N^2 - 1}{N} \cdot\xx = 
\nonumber \\
&=&
 d \lr{2 d - 2}^{-1/2} T^{1/2} \lr{1 - \frac{1}{N^2}} .
\end{eqnarray}

\begin{figure*}
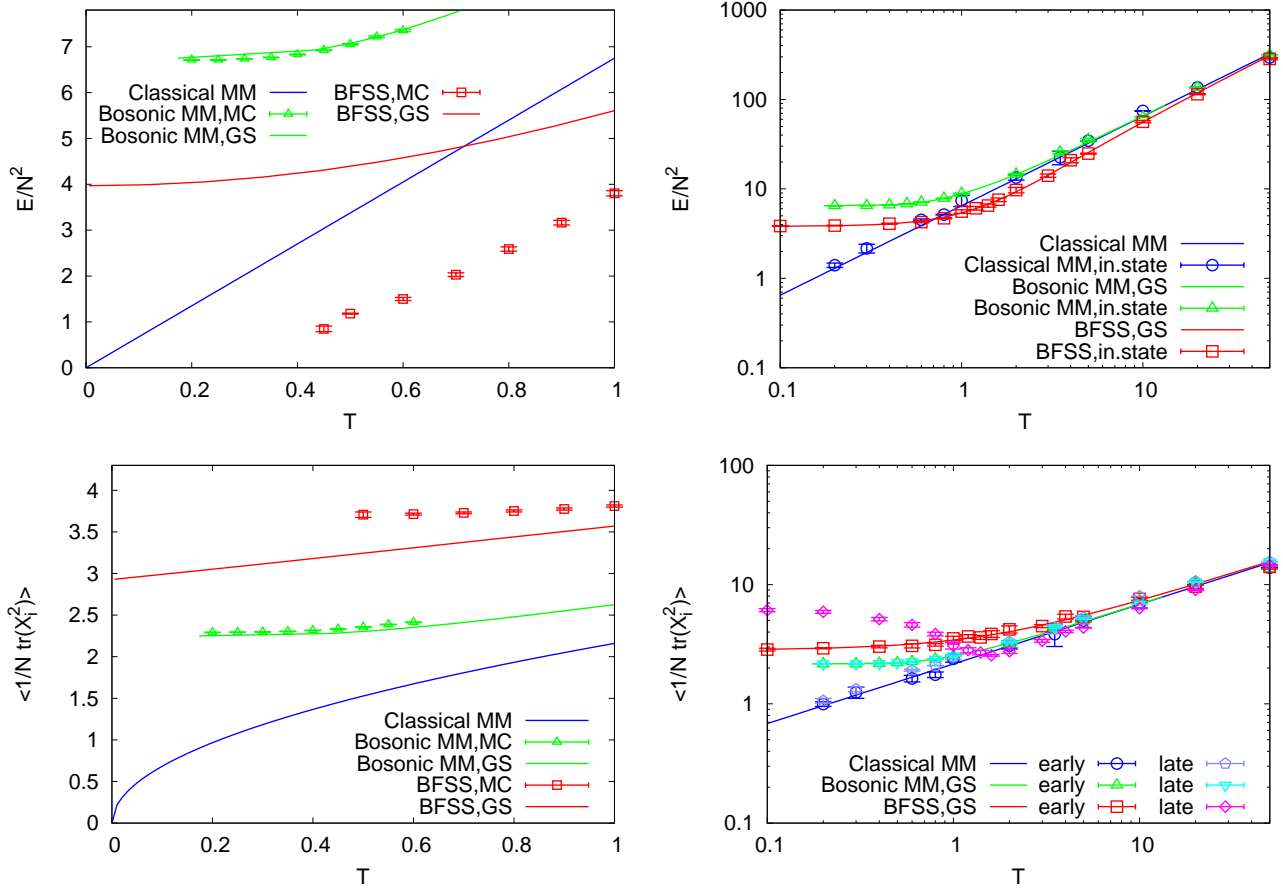

 \centering
 \includegraphics[angle=-90,width=0.48\textwidth]{{{gs_ungauged_eos}}}
 \includegraphics[angle=-90,width=0.48\textwidth]{{{E_vs_T_ourparams}}}\\
 \includegraphics[angle=-90,width=0.48\textwidth]{{{R2_ungauged}}}
 \includegraphics[angle=-90,width=0.48\textwidth]{{{R2_vs_T_ourparams}}}\\
 \caption{Equation of state for the ungauged bosonic matrix model and the ungauged BFSS model within the Gaussian state approximation compared with the results of Monte-Carlo simulations \cite{Hanada:1802.02985} after $N \rightarrow \infty$ extrapolation (plots on the left) and with observables obtained for random initial conditions used in our real-time simulations with $N=5$ (plots on the right). At the top: energy as a function of temperature. At the bottom: coordinate dispersion $\frac{1}{N}\vev{ \tr\hX_i^2}$ as a function of temperature. For this quantity, we present the expectation values in the initial state as well as time-averaged late-time expectation values, which differ significantly in the case of the full BFSS model.}
  \label{fig:gs_eos}
\end{figure*}

In Fig.~\ref{fig:gs_eos} (plots on the left) we compare our equation of state given by equations (\ref{equilibrium_vals_vs_f}) and (\ref{temperature_def}) with the numerical results of \cite{Hanada:1802.02985} for the temperature dependence of the energy and the coordinate dispersion $\frac{1}{N}\vev{ \tr X_i^2} = \frac{1}{N} \vev{\hX^a_i \hX^a_i}$ obtained from first-principle Monte-Carlo simulations of the ungauged bosonic matrix model. The normalization of the coordinate dispersion $\frac{1}{N} \vev{\hX^a_i \hX^a_i}$ is such that it remains finite in the t'Hooft large-$N$ limit.

We indeed observe a rather good agreement within a few percent accuracy for the temperature dependence of both the energy and the coordinate dispersion for all simulation parameters used in \cite{Hanada:1802.02985}. It is also interesting to note that the Gaussian state approximation also reproduces very precisely the prediction of \cite{Maldacena:1802.00428} for the low-temperature behavior of the energy of the ungauged model. Expanding the equations (\ref{temperature_def}) and (\ref{equilibrium_vals_vs_f}) to the leading order in $f - 1/2$, it is easy to obtain the low-temperature asymptotics of the equation of state
\begin{eqnarray}
 E\lr{T} - E\lr{T=0}
 = \nonumber \\ =
 d \lr{d-1}^{1/3} e^{-\frac{\lr{d-1}^{1/3}}{T}}
 \quad (T \ll 1) .
 \label{gs_lowtemp_scaling}
\end{eqnarray}
The coefficients $d \lr{d-1}^{1/3} = 18$ and $\lr{d-1}^{1/3} = 2$ in (\ref{gs_lowtemp_scaling}) match within statistical errors the numerical fit of the form $E\lr{T} - E\lr{T=0} = B e^{-C/T}$ in \cite{Hanada:1802.02985}, which yields $B = 20.0(2.9)$ and $C = 2.043(76)$. Such a good agreement can be probably explained by the fact that in the dual holographic picture the difference $E\lr{T} - E\lr{T=0}$ at $T \ll 1$ is saturated by rather heavy open string excitations, for which the mean-field-like approximation should work rather well. It also suggests that the Gaussian state approximation is not invalidated in the large-$N$ limit. In particular, the ground state energy
\begin{eqnarray}
\label{gs_energy_approximation}
 E_0/N^2
 \equiv
E\lr{T = 0}/N^2
= \nonumber \\ = 
 \frac{3 d \lr{d-1}^{1/3}}{8} \, \lr{1 - \frac{1}{N^2}}
= \nonumber \\ = 
 6.75 \, \lr{1 - \frac{1}{N^2}}
\end{eqnarray}
in the Gaussian state approximation deviates from the large-$N$ extrapolation of the Monte-Carlo results of \cite{Hanada:1802.02985} by $1 \%$ only. For comparison, applying the Gaussian state approximation to the one-dimensional anharmonic oscillator with the Hamiltonian $\hat{H} = \hat{p}^2 + \hat{x}^4$, one obtains the ground state energy with $2 \%$ precision \cite{Turbiner:math-ph/0506033}. The fact that the $D=9+1$-dimensional bosonic matrix model is very well described by the Gaussian approximation (which is equivalent to mean-field approximation) has been previously noticed in \cite{OConnor:1506.01366}, and explained in terms of the leading order of $1/D$ expansion, which very accurately describes the case of $D=9+1$.

There are two possible ways to interpret the Gaussian state characterized by the correlators (\ref{bfss_exact_initial}) with ${\xp = 0}$ and $\xx$ and $\pp$ given by (\ref{equilibrium_vals_vs_f}) as initial conditions for the real-time dynamics described by equations (\ref{gs_onepoint_eqs}) and (\ref{gs_twopoint_eqs}). 

i) \, A trivial way is to directly substitute the correlators (\ref{bfss_exact_initial}) into equations (\ref{gs_onepoint_eqs}) and (\ref{gs_twopoint_eqs}) which govern the real-time evolution. It is straightforward to check that for all values of $f$ this simply yields a time-independent solution with no particularly interesting properties. A conventional stability analysis based on linearization of equations (\ref{gs_onepoint_eqs}) and (\ref{gs_twopoint_eqs}) also shows that this time-independent solution is stable under small perturbations. Up to corrections proportional to $1/N^2$, small oscillations of $X^a_i$ and $P^a_i$ around $X^a_i = 0$, $P^a_i = 0$ have real-valued frequency
\begin{eqnarray}
\label{frequency_smallX}
 w^2_{X} = 2 \lr{d - 1} N \xx = \frac{2 d - 2}{d} \frac{1}{N} \left\langle\tr \hX_i^2\right\rangle .
\end{eqnarray}
Small oscillations of two-point correlators $\cev{\hX^a_i \hX^b_j}$, $\cev{\hX^a_i \hP^b_j}$ and $\cev{\hP^a_i \hP^b_j}$ around the values (\ref{bfss_exact_initial}) have the frequency
\begin{eqnarray}
\label{frequency_smallXX}
 w^2_{XX} = 12 \lr{d - 1} N \xx = 6 \, w^2_{X} .
\end{eqnarray}
In particular, with such an interpretation we cannot extract any nontrivial Lyapunov exponents and also cannot reproduce the known chaotic behavior of the system in the classical limit. Also since $X^a_i = 0$ for this solution, the fermions completely decouple and we cannot capture their influence on real-time dynamics.

ii) \, In what follows we use another, physically better motivated option of interpreting mixed Gaussian states with $f^2 = \xx \pp - \xp^2 > 1/4$. Namely, we represent the coordinate and momentum dispersions
\begin{eqnarray}
\label{dispersion_splitting_def}
 \xx = \xx^0 + \xx^c,
 \quad
 \pp = \pp^0 + \pp^c,
\end{eqnarray}
as sums of the quantum contributions $\xx^0$ and $\pp^0$ which saturate the uncertainty relation $\xx^0 \pp^0 = 1/4$, and the classical contributions $\xx^c $, $\pp^c$ which describe classical thermal fluctuations. For the purely quantum dispersions $\xx^0$ and $\pp^0$ we use the values (\ref{equilibrium_vals_vs_f}) with $f = 1/2$, which correspond to the Gaussian state $\ket{\Psi_0}$ with lowest possible energy (\ref{gs_energy_approximation}). As long as only the total dispersions $\xx = \xx^0 + \xx^c$ and $\pp = \pp^0 + \pp^c$ enter the variational analysis of the equation of state, the choice of $\xx^0$ is ambiguous. While for the sake of simplicity we choose the value of $\xx^0$ which corresponds to the lowest-energy Gaussian state, in principle one can also make $\xx^0$ a temperature-dependent quantity.

We then represent the finite-temperature Gaussian density matrix characterized by correlators (\ref{bfss_exact_initial}) as a mixture of pure Gaussian states
\begin{eqnarray}
\label{shifted_gaussian}
 \ket{X, P} = \expa{i X^a_i \hP^a_i +  i P^a_i \hX^a_i} \ket{\Psi_0}
\end{eqnarray}
with random coordinate and momentum displacements $X^a_i$ and $P^a_i$ which have Gaussian distributions with dispersions $\vev{X^a_i X^b_j}_c = \xx^c \delta_{ab} \delta_{ij}$, $\vev{P^a_i P^b_j}_c = \pp^c \delta_{ab} \delta_{ij}$:
\begin{eqnarray}
\label{thermal_dm_mixture}
 \hat{\rho} = \left\langle \ket{X, P} \bra{X, P} \right \rangle_c ,
\end{eqnarray}
where $\vev{}_c$ denotes averaging over the classical probability distribution. We then use equations (\ref{gs_onepoint_eqs}) and (\ref{gs_twopoint_eqs}) to individually evolve each of the randomly shifted pure states $\ket{X, P}$ in time. Expectation values of physical observables are finally averaged over random initial values of $X^a_i$ and $P^a_i$. As one can see from the upper right plot in Fig.~\ref{fig:gs_eos}, this representation of the initial thermal state of the system yields the correct temperature dependence of the energy with rather small statistical errors. In the lower right plot in Fig.~\ref{fig:gs_eos} we demonstrate that in this way we reproduce also the correct temperature dependence of the coordinate dispersion $\frac{1}{N}\vev{ \tr\hX_i^2}$, which, unlike energy, is not conserved and has no reason to stay constant in time. Nevertheless, we find that both the early-time expectation value as well as the time-averaged late-time expectation values of this observable agree very well with the thermal equation of state. These observations justify the interpretation of a mixture of nontrivial time-dependent pure states as a dynamical equilibrium state. In Fig.~\ref{fig:X2_vs_time} in the next Section~\ref{sec:num_res} we also show the full time dependence of $\frac{1}{N}\vev{ \tr\hX_i^2}$.

This interpretation of the classical component of the dispersions of $X^a_i$ and $P^a_i$ also allows to make contact with classically chaotic behavior at high temperatures. Indeed, at high temperatures the classical dispersions will strongly dominate over the quantum ones, and the dynamics described by equations (\ref{gs_onepoint_eqs}) and (\ref{gs_twopoint_eqs}) becomes very close to the classical one. Due to its chaoticity and ergodicity, classical matrix mechanics exhibits real-time thermalization towards a dynamical equilibrium state in which long-time averages of physical observables approach their thermal equilibrium values \cite{Hanada:1512.00019}. This thermalization process can be also interpreted as quasinormal ringing characterized by nontrivial complex-valued quasinormal frequencies \cite{Hanada:1809.01671,Aprile:1611.00786}, with real parts being quite close to our estimates (\ref{frequency_smallX}) and (\ref{frequency_smallXX}), see Subsection~\ref{subsec:qnfs}.
At the same time, classical matrix mechanics has finite Lyapunov exponents. Thus interpreting thermal states as dynamical equilibrium states we can capture quantum corrections to Lyapunov exponents and imaginary parts of quasinormal frequencies as well as the time evolution of quantum entanglement.

Of course, the two interpretations discussed above would be equivalent for unitary evolution, but yield drastically different results for the non-unitary evolution within the Gaussian state approximation.

\subsection{Full BFSS model}
\label{subsec:eos_fermionic}

To obtain the equation of state of the full BFSS model we will use the same approach as for the bosonic matrix model and find mixed Gaussian states of fixed energy which maximize the von Neumann entropy. To this end we again split the coordinate and momentum dispersions into the classical and quantum contributions, as in (\ref{dispersion_splitting_def}) and (\ref{thermal_dm_mixture}). Since according to equation (\ref{gs_twopoint_eqs_psipsi}) fermionic degrees of freedom only interact with the classical expectation value $X^a_i$, we assume that the fermions are initially in the ground state with fixed classical coordinates $X^a_i$. This assumption is in line with our construction of the thermal initial conditions where the thermal state was represented by a mixture of ground-state wave functions averaged over random wavepacket shifts. Introducing some finite initial temperature for fermions will only increase the energy of our mixed states, which for our approximation is anyway higher than the exact value for the full BFSS model (see topmost left plot on Fig.~\ref{fig:gs_eos}), and thus will not improve our approximation.

Correspondingly, the fermionic contribution $E_F$ to the energy only depends on the dispersion $\xx^c$ of the classical wave-function shifts $X^a_i$ in (\ref{thermal_dm_mixture}), and is obtained by averaging the ground-state energy of the fermionic Hamiltonian over a Gaussian ensemble of classical coordinates $X^a_i$:
\begin{eqnarray}
\label{fermionic_energy_mean_def}
 E_F = \left\langle \frac{i}{2} \, C_{abc}  \, \sigma_i^{\alpha\beta} \, X^b_i \cev{\hpsi^a_{\alpha} \hpsi^c_{\beta}} \right\rangle_c \, .
\end{eqnarray}
A detailed discussion of the spectrum and the ground state of this Hamiltonian is given in Appendix~\ref{apdx:majorana_initial_state}. Since the only energy scale for the fermionic Hamiltonian is set by the classical $X$ coordinates, dimensional analysis implies that the mean energy of fermions in (\ref{fermionic_energy_mean}) should scale as $E_F \sim \sqrt{\xx^c}$. General scaling arguments from random matrix theory fix the scaling of $E_F$ with $N$, which allows to estimate $E_F$ up to an overall universal coefficient as
\begin{eqnarray}
\label{fermionic_energy_mean}
 E_F = - A_f \, \lr{N^2 - 1} \sqrt{N \, \xx^c} .
\end{eqnarray}
While the $N$- and $\xx^c$-independent coefficient $A_f$ can be calculated exactly using the methods of random matrix theory, for the purposes of this work we obtain the value of $A_f$ numerically by averaging the fermionic energy over sufficiently large ensemble of randomly generated $X^a_i$ coordinates and fitting the dependence on $\xx^c$ and $N$ to equation (\ref{fermionic_energy_mean}). These fits work perfectly within statistical errors and yield $A_f = 15.2661(34)$.

Since fermions are assumed to be in the ground state at fixed $X^a_i$, by virtue of Nernst's theorem their contribution to von Neumann entropy is zero. Thus in order to obtain the equation of state of the full BFSS model within the Gaussian state approximation, we have to maximize the von Neumann entropy (\ref{equilibrium_von_Neumann}) at fixed energy
\begin{eqnarray}
\label{equilibrium_energy_vev_ferm}
 E
 &=&
 d \lr{N^2 - 1} \lr{\frac{\pp}{2 N} + \frac{N^2 \xx^2 \lr{d-1}}{2}}
- \nonumber \\
& &
-
 A_f \, \lr{N^2 - 1} \sqrt{N \, \xx^c} .
\end{eqnarray}
Since the fermionic contribution depends only on the dispersion $\xx^c$ of classical coordinates, the entropy should be maximized with respect to both $\xx^c$ and $\xx^0$, which are now considered as independent parameters. In particular, the separation (\ref{dispersion_splitting_def}) of the coordinate dispersion into the quantum and classical contributions is no longer ambiguous.

In order to maximize the von Neumann entropy (\ref{equilibrium_von_Neumann}) in the space of Gaussian states with fixed energy, we now use equation (\ref{equilibrium_energy_vev_ferm}) to express $\pp$ in terms of $\xx^c$, $\xx^0$ and $E$, which allows to express $f^2 = \xx \pp$ in (\ref{equilibrium_von_Neumann}) as
\begin{eqnarray}
\label{f2_vs_sxx}
 f^2
 &=& \xx \pp =
 \nonumber\\
 &=&
 \lr{\xx^0 + \xx^c}
 \left(
  \frac{2 N E}{d \lr{N^2 - 1}}
  +
  \frac{2 N A_f}{d} \sqrt{N \xx^c}
  - \right. \nonumber \\
  & &
  \qquad\left.
  -
  N^3 \lr{\xx^0 + \xx^c}^2 \, \lr{d - 1}
 \right) .
\end{eqnarray}
This function does not have a local minimum in the space of $\xx^0$ and $\xx^c$, that is, the equations $\partial f^2/\partial \xx^0 = 0$ and $\partial f^2/\partial \xx^c = 0$ have no physical solutions. We have to remember, however, that the classical and quantum dispersions have to be all nonnegative and satisfy $\xx^0 \pp^0 = 1/4$. This turns the maximization of (\ref{f2_vs_sxx}) into a constrained optimization problem, for which the extremum might lie on the boundary of a region allowed by constraints. To describe this region we express $\pp^0$ and $\pp^c$ in terms of $f^2$, $\xx^0$ and $\xx^c$ as
\begin{eqnarray}
\label{spp_constraint}
 \pp^0 = \frac{1}{4 \xx^0},
 \quad
 \pp^c = \frac{f^2}{\xx^0 + \xx^c} - \frac{1}{4 \xx^0} .
\end{eqnarray}
Inserting the explicit expression (\ref{f2_vs_sxx}) for $f^2$ into the above formula for $\pp^c$, after some algebra we can rewrite the constraint $\pp^c > 0$ solely in terms of $\xx^0$ and $\xx^c$ as
\begin{eqnarray}
\label{positive_pp_constraint}
\frac{d \lr{d-1} N^2}{2} \lr{\xx^0 + \xx^c}^2
 +
 \frac{d}{8 \xx^0 N}
 - \nonumber\\ - 
 A_f \sqrt{N \xx^c}
 \leq
 \frac{E}{N^2 - 1} .
\end{eqnarray}
The minimal value of the function on the left-hand side of this inequality sets the lowest value of energy at which the constraint can still be satisfied. Numerical minimization yields $E > E_0 = 3.9692 \lr{N^2 - 1}$. Thus $E_0$ is the ground state energy of the full BFSS model within the Gaussian state approximation. While it is noticeably lower than the value (\ref{gs_energy_approximation}) for the bosonic matrix model, supersymmetry of the full BFSS model implies that the true ground state energy should vanish (see e.g. \cite{Hanada:1802.02985,Hanada:0707.4454}). Again we see that supersymmetry cannot be preserved within the Gaussian state approximation (see also Appendix~\ref{apdx:csft_symmetries}). To get a lower ground state energy, one needs to include at least the three-point connected correlators of the form $\cev{\hX^a_i \hpsi^b_{\alpha} \hpsi^c_{\beta}}$ in the numerical analysis.

We now obtain the equation of state for the full BFSS model within the Gaussian state approximation by maximizing the von Neumann entropy (\ref{equilibrium_von_Neumann}) with respect to $\xx^0$ and $\xx^c$ within the region specified by the constraint (\ref{positive_pp_constraint}) and $\xx^0 > 0$ and $\xx^c > 0$. The value of $f$ is now given by (\ref{f2_vs_sxx}). Since $f^2$ given by (\ref{f2_vs_sxx}) has no local maxima, its maximum lies on the boundary of the optimization region, that is, at $\pp^c = 0$. The corresponding constrained optimization problem cannot be solved exactly, and we use numerical maximization. In this way we obtain $\xx^0$, $\xx^c$ and the von Neumann entropy (\ref{equilibrium_von_Neumann}) as functions of energy $E$. Using numerical interpolation, differentiation and functional inversion, we then again use the relation $T^{-1} = \frac{\partial S}{\partial E}$ to introduce the temperature $T$ and express $\xx^0$, $\xx^c$ and $E$ as functions of $T$. The resulting equation of state is illustrated in Fig.~\ref{fig:gs_eos}, where we show the temperature dependence of the energy and the coordinate dispersion $\vev{\frac{1}{N} \tr X_i^2}$. As one can see from the plots on the left in Fig.~\ref{fig:gs_eos}, the agreement with first-principle numerical simulations of \cite{Hanada:1802.02985} is not so good as for the bosonic matrix model. Nevertheless, the Gaussian state approximation correctly captures the following features of the thermal states of the full BFSS model:
\begin{itemize}
 \item The ground-state energy of the BFSS model is smaller than that of the bosonic matrix model
 \item The coordinate dispersion $\vev{\frac{1}{N} \tr X_i^2}$ is larger than for the bosonic model
 \item At high temperatures both energy and $\vev{\frac{1}{N} \tr X_i^2}$ approach their values in the classical matrix model
\end{itemize}
Let us note that broken supersymmetry and the finiteness of the ground-state energy lead to the non-vanishing classical dispersion $\xx^c$ of $X^a_i$ coordinates, so that the ground state remains disordered and has a finite von Neumann entropy. These features should be absent in the full quantum-mechanical treatment of the BFSS model and should be regarded as artifacts of the Gaussian state approximation.

Having obtained the equation of state and $\xx^0$, $\xx^c$, $\pp^0$ and $\pp^c$ as functions of temperature, we again interpret the classical dispersion $\xx^c$ in terms of a mixture (\ref{thermal_dm_mixture}) of pure states with randomly shifted wave functions. Initial values of the fermionic correlators $\cev{\hpsi^a_{\alpha} \hpsi^b_{\beta}}$ are fixed by assuming that the Majorana fermions are in the ground state at fixed coordinate expectation values $X^a_i$, see Appendix~\ref{apdx:majorana_initial_state} for a more detailed discussion. Simulating the real-time evolution of these pure states, we then average the result over random shifts in the initial conditions. In the upper right plot in Fig.~\ref{fig:gs_eos} we demonstrate that such averaging correctly reproduces the temperature dependence of the energy. On the other hand, for the coordinate dispersion $\frac{1}{N}\vev{ \tr\hX_i^2}$ the correct temperature dependence is only reproduced by early-time averages and by late-time averages at sufficiently high temperatures. At low temperatures the late-time averages deviate significantly from their thermal values, which might be related to the conjectured real-time instability of the BFSS model with respect to spontaneous emission of $D0$-branes \cite{Hanada:1602.01473}.

As a side remark, let us note that the temperature-dependent energies, coordinate dispersions and entropies obtained within the Gaussian state approximation satisfy the so-called Bekenstein bound $S \leq 2 \pi E R$ \cite{Bekenstein:PhysRevD.23.287,Page:1804.10623}, with $R$ defined as
$R = \sqrt{\frac{1}{N} \langle\tr\hX_i^2\rangle}$.
In fact, for all the models which we consider (classical matrix mechanics, bosonic matrix model and the full BFSS model) we have $S \ll 2 \pi E R$.

\section{Numerical results}
\label{sec:num_res}

In this work we numerically solve equations (\ref{gs_onepoint_eqs}) and (\ref{gs_twopoint_eqs}) with initial conditions described in Section~\ref{sec:thermal_initial}. We use a discretization scheme described in Appendix~\ref{apdx:discretization}. Since the numerical cost of our simulations scales as $N^5$, we mostly use a moderately large value $N = 5$. We have also performed a few simulations with $N = 7$ to make sure that our results do not change qualitatively at larger $N$ and exhibit the proper t'Hooft scaling.

We average simulation results over several (typically, between five and seven) random initial conditions as previously discussed in Section~\ref{sec:thermal_initial}. Where shown, error bars on our plots represent the statistical error for such an averaging. Since the number of degrees of freedom in our model is sufficiently large, this statistical error is typically very small due to self-averaging, which works well even for a single instance of random initial conditions. In particular, due to high numerical cost we have used only a single instance of random initial conditions for simulations with $N = 7$, thus the corresponding data points on our plots have no error bars.

\begin{figure*}
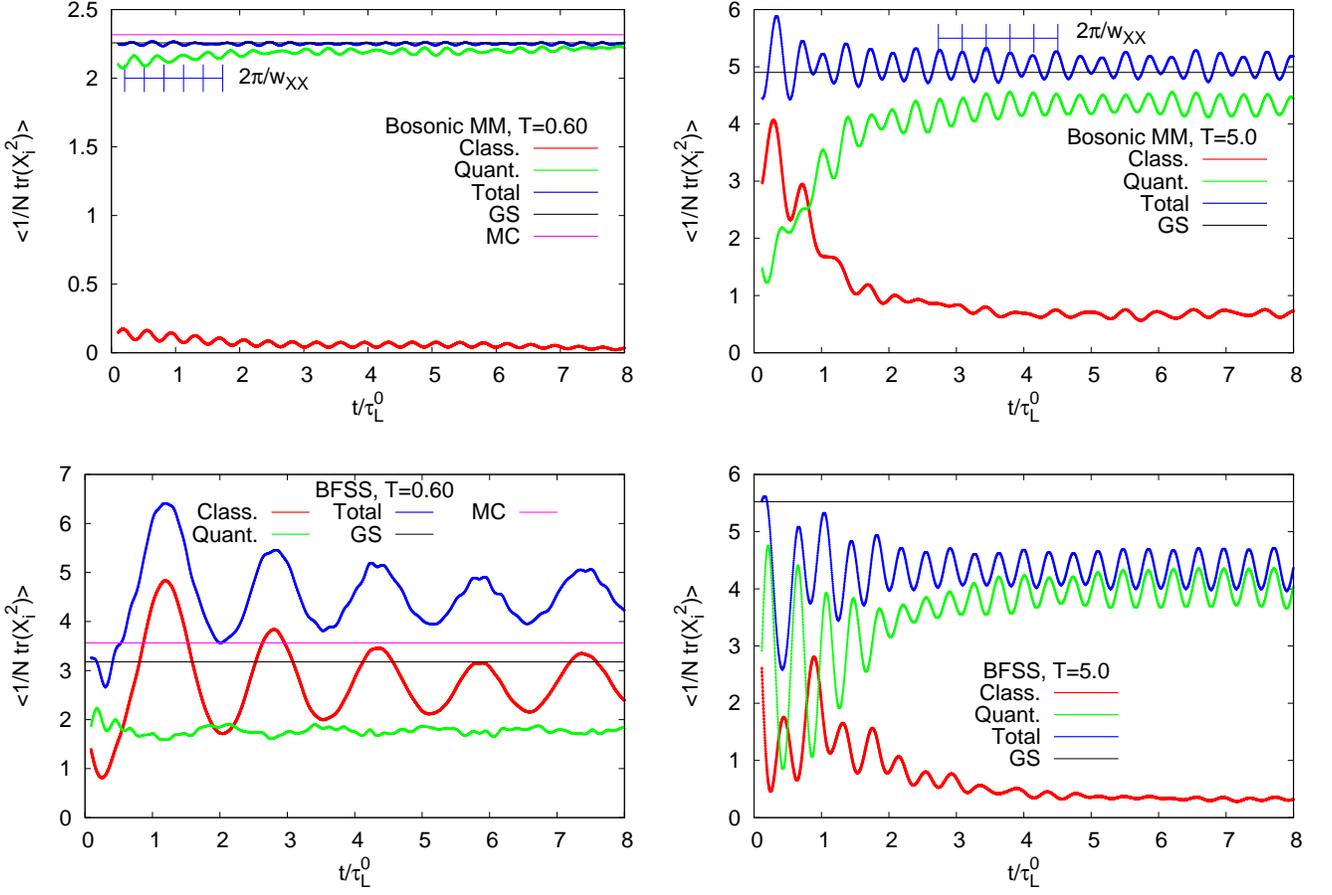

  \centering
  \includegraphics[angle=-90,width=0.49\textwidth]{{{X2_vs_time_bosonic_lowT}}}
  \includegraphics[angle=-90,width=0.49\textwidth]{{{X2_vs_time_bosonic_highT}}}\\
  \includegraphics[angle=-90,width=0.49\textwidth]{{{X2_vs_time_BFSS_lowT}}}
  \includegraphics[angle=-90,width=0.49\textwidth]{{{X2_vs_time_BFSS_highT}}}\\
 \caption{Time dependence of the coordinate dispersion $\frac{1}{N}\vev{ \tr \hX_i^2}$ for $N=5$ in the low-temperature (plots on the left) and high-temperature (plots on the right) regimes of the bosonic matrix model (plots at the top) and the BFSS model (plots at the bottom). We show the classical and the quantum contributions to this quantity along with their total. Solid horizontal lines show the thermal expectation value obtained within the Gaussian state approximation and from first-principle Monte-Carlo simulations of \cite{Hanada:1802.02985}. The gratings show the period $2 \pi/w_{XX}$ of lowest-frequency quasinormal oscillations  of the coordinate dispersion around the thermal state, with $w_{XX}$ given by (\ref{frequency_smallXX}).}
  \label{fig:X2_vs_time}
\end{figure*}

To have a first look at the real-time dynamics described by equations (\ref{gs_onepoint_eqs}) and (\ref{gs_twopoint_eqs}), in Fig.~\ref{fig:X2_vs_time} we show the time dependence of the coordinate dispersion $\frac{1}{N} \langle\tr\hX_i^2\rangle = \frac{1}{N} \langle\hX^a_i \hX^a_i\rangle$ for real-time simulations of the bosonic matrix model and the full BFSS model at different temperatures.

In order to make a meaningful comparison of simulations with characteristic time scales which differ by several orders of magnitude, in Fig.~\ref{fig:X2_vs_time} and in other plots in this work we express physical time in units of the classical Lyapunov time $\tau_L^0$, which is defined as the inverse of the leading Lyapunov exponent $\lambda_L^0$ for the classical dynamics of the BFSS model (\ref{bfss_Hamiltonian}) at a given temperature. For classical matrix mechanics, the temperature dependence of the leading Lyapunov exponent is known to be \cite{Hanada:1512.00019}
\begin{eqnarray}
\label{classic_bfss_lyapunov}
 \lambda_L^0 = \lr{0.292 - \frac{0.42}{N^2}} T^{1/4} .
\end{eqnarray}

In Fig.~\ref{fig:X2_vs_time} we also separately show the contributions of the classical dispersion $\vev{ X^a_i X^b_j}_c$ and the quantum dispersion $\cev{\hX^a_i \hX^b_j}$. Their sum is the physical observable $\frac{1}{N}\langle \tr\hX_i^2\rangle$. For the bosonic matrix model at all temperatures as well as for the BFSS model at high temperatures we observe that within a short time interval $t < 2 \, \tau_L^0$ the classical contribution to the coordinate dispersion tends to decrease by a factor of roughly four. At the same time, the quantum contribution grows in such a way that the total coordinate dispersion remains practically constant in time, up to small short-scale fluctuations with characteristic frequency close to $w_{XX}$ as given by (\ref{frequency_smallXX}).

The decrease of the classical contribution and the corresponding increase of the quantum contribution become particularly large at high temperatures, which indicates a rapid spread of wave functions in configuration space driven by the chaotic dynamics of their centers. In this way the system thus approaches a state of dynamical equilibrium. However, despite this rearrangement, it turns out that the overall coordinate and momentum dispersions which determine the von Neumann entropy (\ref{gs_von_Neumann}) in our simulations remain practically constant in time and exhibit only small short-scale fluctuations. Since the energy is also conserved, we can still assign an approximate value of temperature to this dynamical equilibrium state and use the thermal equations of state derived in the previous Section~\ref{sec:thermal_initial}.

Of course, one should keep in mind that the Gaussian state approximation probably becomes invalid at late times, more precisely, at time of order of several classical Lyapunov times $\tau_L^0$. In particular, at such late times the wave function is already strongly delocalized, and one can expect that non-Gaussian terms in the Schr\"{o}dinger equation become important. E.g. in a simple classically chaotic model with two bosonic degrees of freedom we have observed a good quantitative agreement with the numerical solution of the Schr\"{o}dinger equation only up to $t \leq 2 \tau_L^0$ \cite{Buividovich:17:5}. Thus while the values at which the quantum and the classical contributions saturate at late times might be not completely accurate, the enhancement of the quantum dispersion at early times should be a physical feature. For the high-temperature regime of the full BFSS model and also for larger $N=7$ the situation appears to be very similar, in particular, the time at which the wave function spreading starts is practically the same.

The late-time expectation value of the coordinate dispersion $\frac{1}{N}\vev{\tr\hX_i^2}$ deviates significantly from the thermal equation of state only for the low-temperature regime of the full BFSS model, where our approximation is most likely inaccurate due to the violation of supersymmetry. While we cannot say much about the dynamics of the BFSS model in this regime,
the growth of coordinate dispersion might be related to the real-time instability of the thermal state of the BFSS model with respect to spontaneous emission of $D0$-branes due to repulsive fermionic forces, similar to Hawking radiation \cite{Hanada:1602.01473}.

\subsection{Lyapunov exponents and the MSS bound}
\label{subsec:lyapunov_mss}

In classical mechanics the distance between two infinitely close solutions of the equations of motions can be expressed in terms of the Poisson bracket $\lrc{X^a_i\lr{t}, P^b_j\lr{0}} = \frac{\partial X^a_i\lr{t}}{\partial X^b_j\lr{0}}$, which is replaced by the commutator $\lrs{\hX^a_i\lr{t}, \hP^b_j\lr{0}}$ in quantum theory. Since for most parity- and time-reversal-invariant thermal density matrices the expectation value of this commutator vanishes, one typically considers expectation values of the form (\ref{otoc_def}) which involve squared commutators \cite{Larkin:JETP1969,Stanford:1304.6483,Maldacena:1503.01409}. In this work we consider the out-of-time-order correlator
\begin{eqnarray}
\label{otoc_our}
 C\lr{t} = \tr \lr{ \hat{\rho} \, \lrs{\hX^a_i\lr{t}, \hP^b_j\lr{0}}^2 } ,
\end{eqnarray}
where $\hat{\rho}$ is the initial Gaussian density matrix at $t = 0$. This definition is a direct generalization of the Lyapunov distance in $X$ space to quantum theory. A generalization of the out-of-time-order correlator (\ref{otoc_our}) can be used to define the full spectrum of Lyapunov exponents for quantum systems \cite{Hanada:1809.01671}.

In order to treat the out-of-time-order correlator (\ref{otoc_our}) within the Gaussian state approximation, we interpret our thermal Gaussian density matrix as a mixture of pure Gaussian states $\ket{X, P}$ with randomly distributed classical expectation values $X^a_i$ and $P^a_i$, as discussed in the previous Section~\ref{sec:thermal_initial}. We further interleave the two commutators in (\ref{otoc_our}) with the identity decomposition $\hat{I} =  \int dX' \, dP' \ket{X',P'} \bra{X',P'}$ in terms of the Gaussian states $\ket{X',P'}$ shifted in coordinate and momentum space by $X'$ and $P'$, which leads to
\begin{eqnarray}
\label{otoc_vs_single_commutator}
 \tr \lr{ \hat{\rho} \, \lrs{\hX^a_i\lr{t}, \hP^b_j\lr{0}}^2 }
 = \nonumber \\ =
 \int dX' \, dP'
 \left \langle \bra{X, P} \lrs{\hX^a_i\lr{t}, \hP^b_j\lr{0}} \ket{X',P'}
 \times \right. \nonumber\\ \left. \times
 \bra{X', P'} \lrs{\hX^a_i\lr{t}, \hP^b_j\lr{0}} \ket{X, P} \right \rangle_c.
\end{eqnarray}
We then represent each of the commutators in terms of the infinitesimal displacement operators at $t = 0$:
\begin{eqnarray}
\label{otoc_vs_single_commutator1}
 \bra{X, P} \lrs{\hX^a_i\lr{t}, \hP^b_j\lr{0}} \ket{X',P'}
 = \nonumber \\ =
 -i \frac{\partial}{\partial \epsilon^b_j} \bra{X, P} e^{i \epsilon \hP\lr{0} } \hX^a_i\lr{t} e^{- i \epsilon \hP\lr{0} } \ket{X',P'} ,
\end{eqnarray}
and similarly for the second commutator in (\ref{otoc_vs_single_commutator}). Calculating the expression for the matrix elements of $X$ between the two Gaussian states and inserting it into (\ref{otoc_vs_single_commutator1}) and (\ref{otoc_vs_single_commutator}), one can show that within the Gaussian state approximation the integral over $X'$ and $P'$ is saturated by the saddle point at $X' = X\lr{0}$, $P' = P\lr{0}$, at which the integrand is just the square of the expectation value of the single commutator $\bra{X, P} \lrs{\hX^a_i\lr{t}, \hP^b_j\lr{0}} \ket{X, P}$. Hence we can read off the quantum corrections to Lyapunov exponents from the $X$- and $P$-averaged norm of the Lyapunov distance vector $\delta X^a_i$ (\ref{our_lyapunov_def}) between the $X^a_i$ coordinates for the two solutions of equations (\ref{gs_onepoint_eqs}) and (\ref{gs_twopoint_eqs}) with initial conditions which differ by an infinitely small coordinate shift $\epsilon$:
\begin{eqnarray}
\label{our_lyapunov_def}
 \delta X^a_i
 = 
 \bra{X, P}
 e^{i \epsilon^b_j \hP^b_j\lr{0} } \hX^a_i\lr{t} e^{- i \epsilon^b_j \hP^b_j\lr{0} }
 \ket{X, P}
 \myeqbreak{-}
  \bra{X, P} \hX^a_i \lr{t} \ket{X, P}
 \myeqbreak{=}
 i \epsilon^b_j \bra{X, P} \lrs{\hP^b_j\lr{0}, \hX^a_i\lr{t} } \ket{X, P} .
\end{eqnarray}
Technically this distance is much easier to calculate than the squared commutator (\ref{otoc_def}).

\begin{figure*}
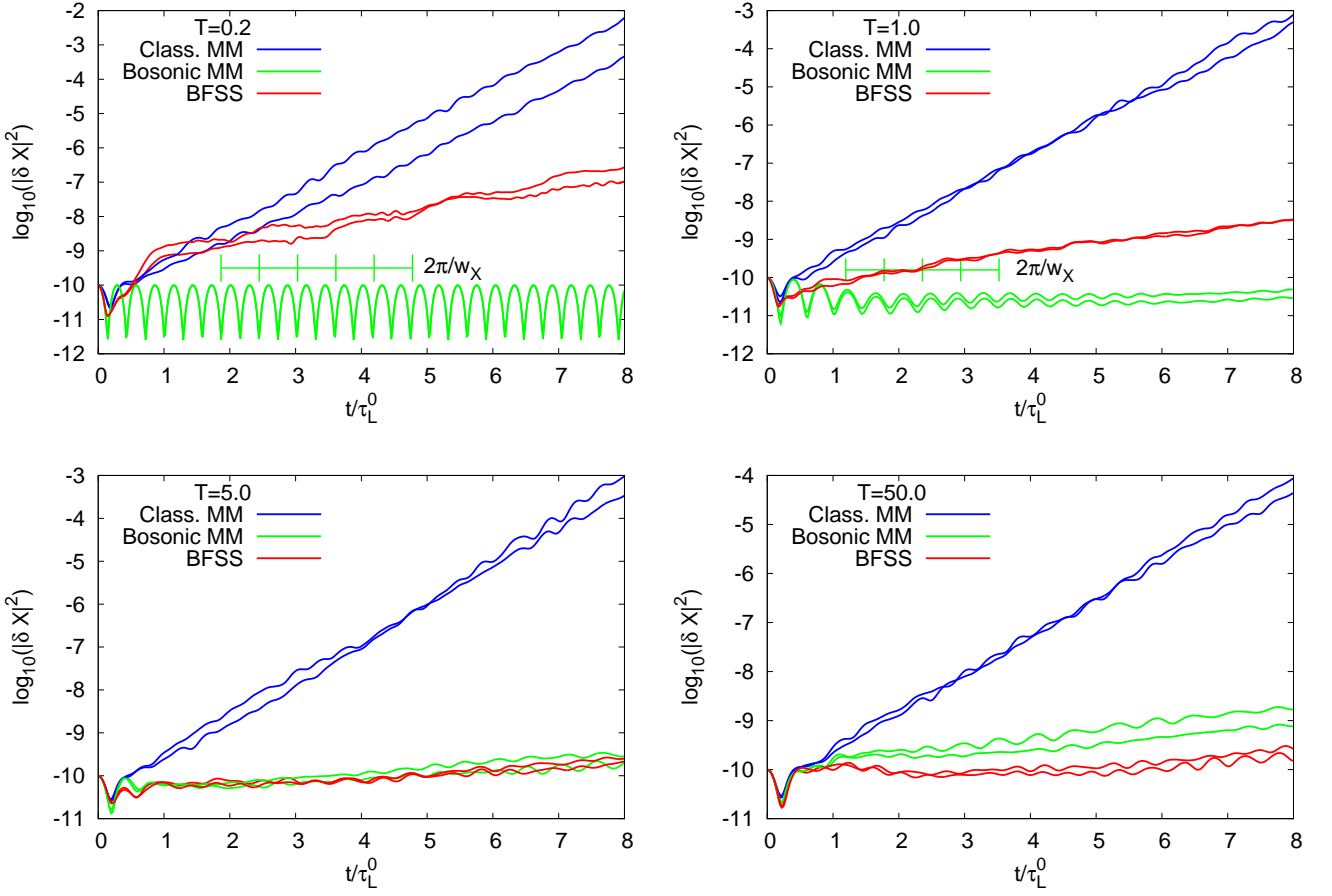

 \centering
 \includegraphics[angle=-90,width=0.49\textwidth]{{{lyapunov_dist_T0.2}}}
 \includegraphics[angle=-90,width=0.49\textwidth]{{{lyapunov_dist_T1.0}}}\\
 \includegraphics[angle=-90,width=0.49\textwidth]{{{lyapunov_dist_T5.0}}}
 \includegraphics[angle=-90,width=0.49\textwidth]{{{lyapunov_dist_T50.0}}}\\
 \caption{Time dependence of the Lyapunov distances $|\delta X^a_i|^2$ for the classical matrix mechanics, bosonic matrix model and the full BFSS model with $N=5$ at different temperatures. We plot the results for a single instance of random initial conditions. The gratings show the period of lowest-frequency oscillations $2 \pi/w_X$ of classical coordinates $X^a_i$, $P^a_i$ around the thermal state, with $w_X$ given by (\ref{frequency_smallX}).}
 \label{fig:lyapunov_dist}
\end{figure*}

In order to calculate the Lyapunov distance as a function of time, we use the first equation in (\ref{our_lyapunov_def}) and consider the distance $|\delta X^a_i|^2$ between two solutions of equations (\ref{gs_onepoint_eqs}) and (\ref{gs_twopoint_eqs}) for which the initial values of the classical expectation values $X^a_i$ differ by a small random vector $\epsilon^a_i$ with $|\epsilon^a_i| = 10^{-5}$. To ensure that the Lyapunov distances grow isotropically in configuration space, we consider at least two different vectors $\epsilon^a_i$. In all of our simulations, Lyapunov distances were growing at the same rate independently of the choice of $\epsilon^a_i$. In Fig.~\ref{fig:lyapunov_dist} we show the time dependence of the Lyapunov distances at different temperatures for $N=5$, comparing classical dynamics with the real-time dynamics of both the bosonic matrix model and the full BFSS model.

While the classical dynamics is always chaotic and exhibits a well-defined exponential growth of the Lyapunov distance, quantum corrections from the bosonic sector make the dynamics completely non-chaotic at sufficiently low temperatures, with Lyapunov distances which do not exhibit any growth, but rather oscillate with a characteristic frequency $w_X$ given by (\ref{frequency_smallX}) which can be obtained by linearizing equations (\ref{gs_onepoint_eqs}) in the vicinity of thermal time-independent solution (\ref{equilibrium_vals_vs_f}). This is an expected non-chaotic behavior in the low-temperature regime of the ungauged bosonic matrix model, where gauging becomes unimportant. Correspondingly, all observables approach their values in the conventional gauged bosonic matrix model, for which the physics in the low-temperature confinement phase does not depend on temperature by virtue of the large-$N$ volume reduction \cite{Nishimura:0706.3517,Aharony:hep-th/0406210}. Since Lyapunov exponents vanish at zero temperature, at any temperature in the confinement regime of the bosonic matrix model Lyapunov exponents should also be zero up to $1/N$ corrections. While there is no strict notion of confinement in the ungauged matrix model, one can still expect an exponential suppression of Lyapunov exponents at low temperatures, by analogy with the low-temperature scaling of energy in (\ref{gs_lowtemp_scaling}).

In contrast, for the BFSS model Lyapunov exponents remain finite down to the lowest temperature which we consider, and Lyapunov distances exhibit a clear growth. At temperatures up to $T \sim 1$ they follow very closely the classical Lyapunov distances in the early evolution period with $t \lesssim 2 \tau_L^0$. At later times the growth becomes milder but is still noticeable. This behavior is in qualitative agreement with the absence of a confinement (or confinement-like) regime in the full BFSS model \cite{Hanada:0707.4454,Hanada:1802.02985}, which is thus expected to be chaotic at all temperatures.

For higher temperatures Lyapunov distances for the bosonic matrix model and the full BFSS model behave in very similar ways. At early times $t \lesssim 0.5 \tau_L^0$ they follow rather precisely the time dependence of the classical Lyapunov distance. Afterwards they exhibit a rather slow exponential growth at a rate which is noticeably smaller than the classical Lyapunov exponent.

We also note that at sufficiently early times, when the exponential growth of Lyapunov distances has not yet fully developed, $|\delta X^a_i|^2$ exhibits signatures of quasinormal ringing, which we will discuss in more details in Subsection~\ref{subsec:qnfs}.

\begin{figure}
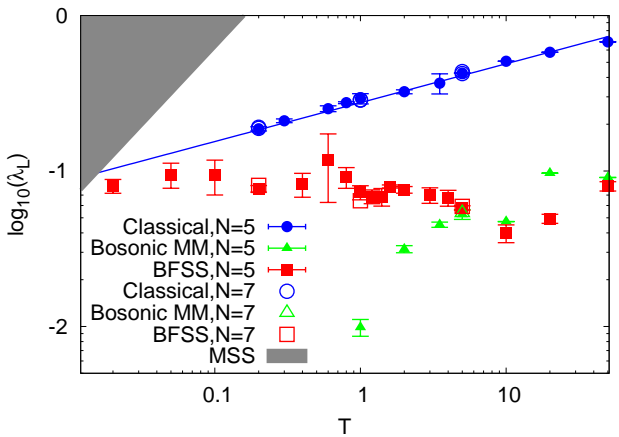

  \centering
  \includegraphics[angle=-90,width=0.48\textwidth]{{{lyapunov_vs_T}}}
  \caption{Leading Lyapunov exponent as a function of temperature for the classical and quantum dynamics of the bosonic matrix model and the full BFSS model for $N = 5$ (filled symbols) and $N = 7$ (empty symbols). We also sketch the MSS bound $\lambda_L < 2 \pi T$ and the scaling law (\ref{classic_bfss_lyapunov}) of the classical Lyapunov exponent (solid blue line).}
  \label{fig:lyapunov_vs_T}
\end{figure}

To summarize, there are no indications that either in the bosonic matrix model or the BFSS model quantum corrections lead to faster growth of Lyapunov distance than in the classical matrix mechanics. At least at sufficiently short evolution times this observation should be qualitatively accurate. To quantify all these observations, we extract the leading quantum Lyapunov exponents $\lambda_L$ by fitting the numerical data for $\ln\lr{|\delta X^a_i\lr{t}|}$ with a linear function of the form $c_0 + \lambda_L t$. In these fits we disregard the transition regime at early times $t/\tau_L^0 \lesssim 1$, where Lyapunov distances between the centers of two infinitesimally close wavepackets are close to the classical Lyapunov distances but do not yet exhibit a clear exponential growth. Obviously, the inclusion of the early evolution period with $t < \tau_L^0$ into the fitting would make our estimates of Lyapunov exponents closer to, but definitely not larger than, the results for the classical matrix mechanics. In particular for the low-temperature regime of the bosonic matrix model the absence of exponential growth is obvious independent of the fitting range. We thus restrict the fitting range to $1 \leq  t/\tau_L^0 \leq 8$, when the exponential growth has already set in and the fits have good quality. While the Gaussian approximation should receive corrections here, especially toward larger $t$, qualitative features -- e.g., that the Lyapunov exponent decreases due to the quantum effect -- should be correct. The temperature dependence of the leading Lyapunov exponents extracted from these fits is illustrated in Fig.~\ref{fig:lyapunov_vs_T} for $N = 5$ and $N = 7$. A comparison of data points for $N = 5$ and $N = 7$ indicates a proper t'Hooft scaling of Lyapunov exponents.

The temperature dependence $\lambda_L^0 \sim T^{1/4}$ of the classical Lyapunov exponent immediately suggests that its value becomes incompatible with the MSS bound $\lambda_L < 2 \pi T$ at the temperature $T^{\star} = \lr{\frac{0.292 - 0.42/N^2}{2 \pi}}^{4/3} = 0.015$, as one can also see from Fig.~\ref{fig:lyapunov_vs_T}. As discussed already in the seminal paper \cite{Maldacena:1503.01409}, this is not a contradiction, since at such low temperatures the classical approximation inevitably breaks down, and quantum effects become important. Our real-time simulations explicitly illustrate this transition between the classical and the quantum regimes. From Fig.~\ref{fig:lyapunov_vs_T} we see that at least for the bosonic matrix model quantum corrections indeed decrease the Lyapunov exponents in such a way that they remain well below the MSS bound at all temperatures.

In the full BFSS model the effect of fermions is to remove the low-temperature confinement-like regime, so that the system remains in the deconfinement phase all the way down to zero temperature independently of gauging \cite{Maldacena:hep-th/9802042,Hanada:0707.4454,Hanada:1802.02985}. Correspondingly, the system should remain chaotic at all temperatures, and the Lyapunov exponents should also remain finite. In agreement with these expectations, for the full BFSS Hamiltonian the leading Lyapunov exponent is always finite in our real-time simulations. In particular, in the low-temperature regime it is significantly larger than the corresponding value for the bosonic matrix model, and tends to approach the corresponding classical value. As we have already discussed, at low temperatures our description of the full BFSS model is probably not accurate enough due to explicitly broken supersymmetry, and we cannot make any strong statements about the validity of the MSS bound. We can only state that our results are compatible with the possibility that the full BFSS model saturates the MSS bound at very low temperatures, similarly to the SYK model.

\subsection{Entanglement entropy generation}
\label{subsec:entanglement_entropy}

Quantum entanglement between different degrees of freedom in an interacting system provides a quantitative picture of the ``scrambling'' and spreading of quantum information. Entanglement can be quantified in terms of the entanglement entropy
\begin{eqnarray}
\label{entanglement_entropy_def}
 S_A = - \tr\lr{\rho_A \ln\lr{\rho_A}},
 \quad
 \rho_A = \tr_B \ket{\Psi} \bra{\Psi} ,
\end{eqnarray}
where $\ket{\Psi}$ is some pure state characterizing the entire system and $A$ and $B$ are the two complementary sets of degrees of freedom which define the decomposition of the Hilbert space $\mathcal{H}$ of the system into a direct product $\mathcal{H} = \mathcal{H}_A \otimes \mathcal{H}_B$. Quantum-chaotic systems are expected to ``scramble'' the information contained in the two subsystems by rapidly entangling the states in $\mathcal{H}_A$ and $\mathcal{H}_B$, whereupon the entanglement entropy quickly reaches some maximal saturation value. For finite-dimensional Hilbert spaces this maximal value is the ``Haar-scrambled'' entanglement entropy \cite{Page:gr-qc/9305007,Susskind:0808.2096}.

Strictly speaking, in gauge theories (of which the BFSS model is a descendant) the splitting of the physical Hilbert space into a direct product $\mathcal{H}_A \otimes \mathcal{H}_B$ is not completely trivial due to gauge constraints \cite{Buividovich:08:3,Donnelly:1109.0036}. This problem, however, is not relevant for us since we work in the ungauged theory \cite{Maldacena:1802.00428} which does not impose the gauge constraints on its Hilbert space by definition.

Numerical calculation of the entanglement entropy is typically a rather nontrivial task, especially for real-time evolution of interacting systems. Since the Gaussian state approximation which we use in this paper evolves pure states into pure states (see Appendix~\ref{apdx:symplectic_conservation}), it also provides a convenient framework for studying quantum entanglement. Entanglement entropy for Gaussian states can be directly calculated in terms of equal-time correlators (\ref{correlators_def}) of canonical variables \cite{Sorkin:PRD1986,Sorkin:1205.2953,Sorkin:1311.7146,Berenstein:1503.04857,Berges:1712.09362}. Here the basic observation is that tracing out degrees of freedom in subsystem $B$ from the Gaussian density matrix $\ket{\Psi} \bra{\Psi}$, as in (\ref{entanglement_entropy_def}), again yields a Gaussian density matrix which is characterized by the same correlators (\ref{correlators_def}), but restricted to canonical variables in subsystem $A$ which describe $N_{dof} \leq N_{tot}$ degrees of freedom. The correlator block matrix of the form (\ref{correlator_block_matrix}) which corresponds to the reduced density matrix $\hat{\rho}_A$ in (\ref{entanglement_entropy_def}) is thus obtained from the full correlator matrix by removing the rows and columns which correspond to degrees of freedom in subsystem $B$ which are being traced out in (\ref{entanglement_entropy_def}) and hence has the size $2 N_{dof} \times 2 N_{dof}$. Being restricted to subsystem $A$, the correlators (\ref{correlators_def}) in general describe a mixed state with a nonzero von Neumann entropy (\ref{gs_von_Neumann}), which is nothing but the entanglement entropy (\ref{entanglement_entropy_def}). Since in our setup the bosonic and fermionic degrees of freedom communicate only via the classical expectation value $X^a_i$, they cannot be entangled quantum mechanically, and we only consider the entanglement between the bosonic degrees of freedom, thus completely tracing out the fermionic Hilbert space.

\begin{figure*}
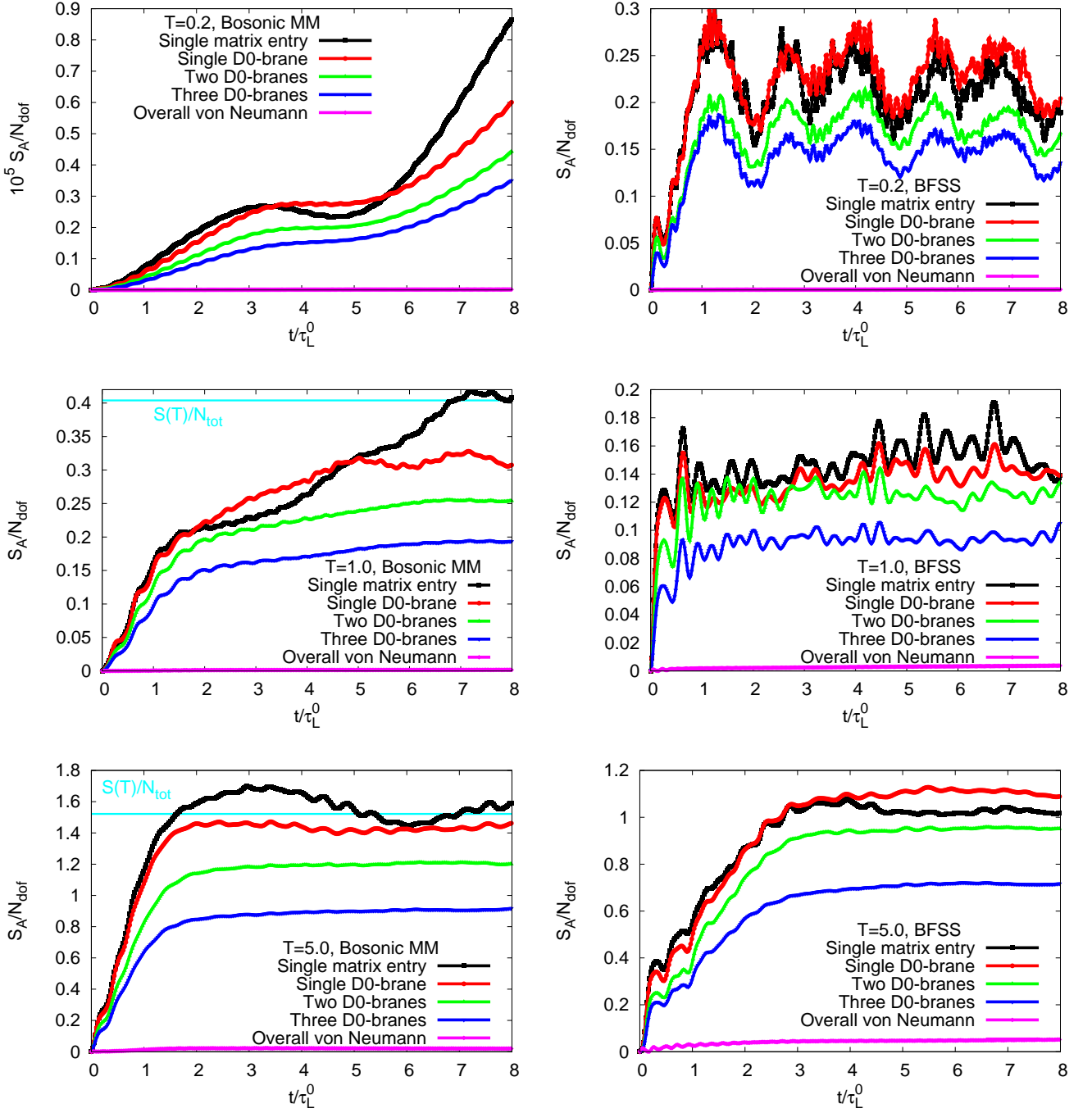

 \centering
 \includegraphics[angle=-90,width=0.49\textwidth]{{{ee_vs_time_bosonic_T0.2}}}
 \includegraphics[angle=-90,width=0.49\textwidth]{{{ee_vs_time_bfss_T0.2}}}\\
 \includegraphics[angle=-90,width=0.49\textwidth]{{{ee_vs_time_bosonic_T1.0}}} \includegraphics[angle=-90,width=0.49\textwidth]{{{ee_vs_time_bfss_T1.0}}}\\
 \includegraphics[angle=-90,width=0.49\textwidth]{{{ee_vs_time_bosonic_T5.0}}}
 \includegraphics[angle=-90,width=0.49\textwidth]{{{ee_vs_time_bfss_T5.0}}}\\
 \caption{Time dependence of the entanglement entropy for different partitions of the Hilbert space in the bosonic matrix model (left) and in the full BFSS model (right) with $N=5$ at different temperatures. Solid horizontal cyan lines correspond to the thermal entropy (\ref{equilibrium_von_Neumann}) divided by the total number of degrees of freedom $N_{tot} = d \lr{N^2 - 1}$ in the system. The deviation of the von Neumann entropy from zero provides us with the estimate of numerical artifacts due to time discretization.}
 \label{fig:ee_vs_time}
\end{figure*}

\begin{figure}
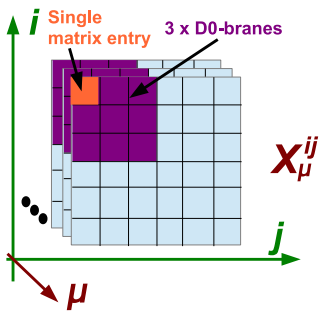

    \centering
    \includegraphics[width=0.25\textwidth]{{{entanglement_illustration}}}
    \caption{Splitting of the matrix entries of $X^{\mu}_{ij}$ into the subsystems $A$ and $B$ for the calculation of the entanglement entropy.}
    \label{fig:entanglement_illustration}
\end{figure}

We now use the prescription sketched above to calculate entanglement entropy for time-dependent correlators (\ref{correlators_def}) obtained by solving equations (\ref{gs_onepoint_eqs}) and (\ref{gs_twopoint_eqs}). In order to make sure that the time dependence of entanglement entropy exhibits universal features independently of subsystem size, we consider four different choices of the subsystem $A$ in (\ref{entanglement_entropy_def}):
\begin{itemize}
 \item \textbf{Single matrix entry with $N_{dof} = 1$:} the indices in the correlator matrix (\ref{correlator_block_matrix}) take values $a = b = a_0$, $i = j = i_0$, where $a_0$ belongs to the Cartan subalgebra of $su\lr{N}$.
 \item \textbf{Single $D0$-brane with $N_{dof} = 9$:} the indices $a = b = a_0$ in (\ref{correlator_block_matrix}) are fixed, while the spatial indices $i$, $j$ take all possible values. We choose $a_0$ to belong to the Cartan subalgebra of $su\lr{N}$, in which case $X^{a_0}_i$ can be interpreted as coordinates of a single $D0$-brane.
 \item \textbf{Two $D0$-branes with $N_{dof} = 36$,} in which case the indices $a$, $b$ in (\ref{correlator_block_matrix}) take two possible values each, and the spatial indices $i$, $j$ take all possible values. Two values $a_0, a_1$ correspond to Cartan matrices for which the maximal matrix entries are at positions $a_0 = \lr{0, 0}$ and $a_1 = \lr{1, 1}$, and two other values correspond to matrices with non-zero elements at off-diagonal positions $a_2 = \lr{1, 0}$ and $a_3 = \lr{0, 1}$. $X^{a_0}_i$ and $X^{a_1}_i$ can be interpreted as the spatial coordinates of two $D0$-branes, and $X^{a_2}_i$ and $X^{a_3}_i$ parameterize the excitations of strings which join these branes \cite{Witten:hep-th/9510135,Susskind:hep-th/9610043}.
\item \textbf{Three $D0$-branes with $N_{dof} = 81$:} this case is analogous to the above case of two $D0$-branes, but the indices $a, b$ take three values each. This gives nine values in total, out of which three values belong to the Cartan subalgebra, and six values correspond to off-diagonal terms of $X_i$ and $P_i$ matrices.
\end{itemize}
These ways of separating the bosonic degrees of freedom of the BFSS model (\ref{bfss_Hamiltonian}) into subsystems $A$ and $B$ are schematically illustrated on Fig.~\ref{fig:entanglement_illustration}.

The splitting of matrix entries of $X_i$ coordinates into diagonal entries which are interpreted as $D0$-brane coordinates and off-diagonal entries which are interpreted as stringy excitations would be particularly obvious for a $u\lr{N}$ Lie algebra, for which the simplest basis of the Cartan subalgebra consists of diagonal matrices with only one unit element. The center of $u\lr{N}$ algebra, proportional to the unit matrix, corresponds to the center of mass of the system which we set to zero in order to exclude a trivial flat direction which corresponds to the overall motion of the center of mass of the system. This leads to an $su\lr{N}$ algebra, for which this splitting is not so straightforward at finite $N$, as orthonormal Cartan matrices have several non-zero elements and thus correspond to displacements of many $D0$-branes simultaneously. Here we simply rely on the fact that at sufficiently large $N$ the difference between $su\lr{N}$ and $u\lr{N}$ becomes negligible, and all but one elements of Cartan matrices are suppressed as $1/N$. We will also see that the time dependence of entanglement entropy shows universal features which are independent of a particular splitting of the degrees of freedom between $A$ and $B$.

We calculate the entanglement entropy separately for each random initial condition, that is, separately for each pure Gaussian state $\ket{X, P}$ in the decomposition (\ref{thermal_dm_mixture}) of the thermal density matrix. For different typical random initial conditions the time dependence of the entanglement entropy appears to be very similar due to self-averaging at sufficiently large $N$. Thus if we choose to calculate entanglement entropy for the mixed density matrix (\ref{thermal_dm_mixture}) averaged over various Gaussian states, we simply obtain a practically time-independent constant contribution to the entanglement entropy which is proportional to $N_{dof}$.
We illustrate the time dependence of the entanglement entropy for $N = 5$ and three different temperatures in Fig.~\ref{fig:ee_vs_time}, normalizing it by the corresponding number of degrees of freedom $N_{dof}$. Plots on the left and on the right correspond to the bosonic matrix model and the full BFSS model, respectively. We plot the data for some particular random initial conditions without any averaging. In order to control numerical errors due to time discretization, in Fig.~\ref{fig:ee_vs_time} we also plot the overall von Neumann entropy of our pure Gaussian state (also divided by $N_{dof}$), which should be zero for continuous time evolution of pure Gaussian states, as discussed in Appendix~\ref{apdx:symplectic_conservation}. The deviation of the von Neumann entropy from zero thus reflects numerical artifacts due to time discretization. It is considerably smaller than the entanglement entropy for all our simulation parameters.

We observe that the entanglement entropy indeed exhibits an expected universal ``scrambling'' behavior: a roughly linear growth at early times and saturation at late times. Only for the low-temperature regime of the bosonic matrix model the growth appears to be so slow that we don't see the onset of saturation up to $t = 8 \tau_L^0$. It also turns out that if the number of degrees of freedom $N_{dof}$ in the entangled subsystem is much smaller than the total number of degrees of freedom $N_{tot}$, entanglement entropy is approximately proportional to $N_{dof}$ at all times, as could be expected for a thermal entropy.

For the bosonic matrix model the saturation value of the ratio $S_A/N_{dof}$ at $N_{dof} \ll N_{tot}$ turns out to be quite close to the von Neumann entropy $S\lr{T}/N_{tot}$ per degree of freedom for the thermal Gaussian state given by (\ref{gs_von_Neumann}). For illustration, we show the value of $S\lr{T}/N_{tot}$ on the left plots in Fig.~\ref{fig:ee_vs_time} with cyan horizontal line.

The observation that \emph{entanglement entropy} per degree of freedom for a single pure state is close to the \emph{thermal von Neumann entropy} per degree of freedom illustrates a nontrivial relation between real-time thermalization of pure states of a sufficiently large system and the description of thermal states in terms of mixed density matrices, as discussed in \cite{Susskind:0808.2096}. This relation is a quantum analogue of the equivalence of micro-canonical and canonical ensembles for chaotic systems, and provides yet another argument in favor of real-time thermalization in our simulations. In this respect entanglement entropy is a convenient measure of the complexity of a pure state, similarly to e.g. Husimi-Wehrl entropy \cite{Schafer:0809.4831}.

Approximately linear scaling of entanglement entropy with the number of degrees of freedom $N_{dof}$ is also in agreement with the findings of \cite{Berenstein:1503.04857}, where it was demonstrated that for periodically driven harmonic oscillators the entanglement generation rate is proportional to the sum of $N_{dof}$ largest Lyapunov exponents. For sufficiently large system with dense Lyapunov spectrum and for sufficiently small $N_{dof}$ this sum should also scale linearly with $N_{dof}$.

When the number of degrees of freedom becomes comparable with the maximal value $N_{tot}$, the ratio $S_A/N_{dof}$ becomes smaller. This behavior is again expected from the identification $N_{dof} \leftrightarrow N_{tot} - N_{dof}$ which follows from the equality $S_A = S_B$. When $N_{dof}$ is small, the entanglement entropy exhibits noticeable fluctuations which are absent at larger $N_{dof}$ due to self-averaging. For bosonic matrix model the typical time during which the entanglement entropy reaches saturation is between one and three classical Lyapunov times at $T \gtrsim 1$.

In the full BFSS model (plots on the right side of Fig.~\ref{fig:ee_vs_time}) fermions completely change the dynamics at low temperatures and make the time at which the entanglement entropy saturates significantly shorter than for the bosonic matrix model at the same temperature. For the full BFSS model the thermal entropy per degree of freedom appears to be around $50-100\%$ larger than the saturation value of the entanglement entropy per degree of freedom, which is probably related to the fact that our approximation is in general worse for the BFSS model than for the bosonic matrix model. At higher temperatures the effect of fermions gradually becomes smaller, and at $T = 5.0$ the time evolution of entanglement entropy is already very similar in both the bosonic matrix model and the full BFSS model.

\begin{figure}
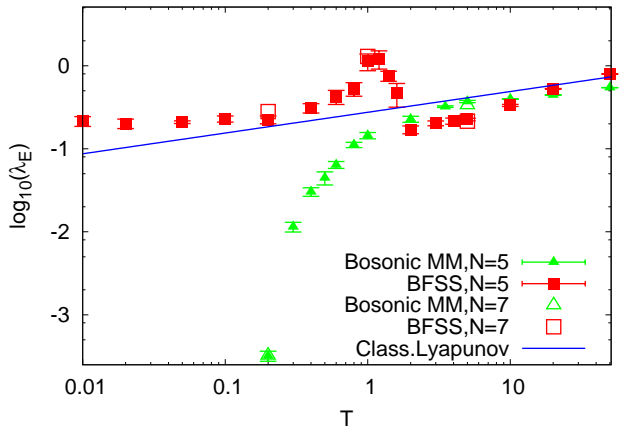

 \centering
 \includegraphics[angle=-90,width=0.48\textwidth]{{{entanglement_vs_T}}}
 \caption{Inverse entanglement saturation time $\lambda_E \equiv \tau_E^{-1}$ as a function of temperature for the bosonic matrix model and the full BFSS model for $N = 5$ (filled symbols) and $N = 7$ (empty symbols). For comparison we also show the temperature dependence of the classical Lyapunov exponent $\lambda_L^0$ given by (\ref{classic_bfss_lyapunov}).}
 \label{fig:entanglement_vs_T}
\end{figure}

In order to quantify the time scale for the saturation of the entanglement entropy more precisely, we define the entanglement saturation time $\tau_E$ by fitting the time-dependent entanglement entropy with a function $A \, \tanh\lr{t/\tau_E}$. The overall normalization constant $A$ in this fit takes care of the subsystem-dependent late-time saturation value of the entanglement entropy, and allows for consistent comparison of $\tau_E$ between simulation results obtained for different temperatures, $N_{dof}$ and $N$. $\tau_E$ also sets the characteristic scale for the entropy production rate:
\begin{eqnarray}
 dS/dt \approx \frac{S\lr{T}}{\tau_E} \, \frac{N_{dof}}{N_{tot}} ,
\end{eqnarray}
where $S\lr{T}$ is the thermal entropy (\ref{gs_von_Neumann}).

By analogy with Lyapunov exponents, we introduce the inverse entanglement saturation time $\lambda_E \equiv \tau_E^{-1}$. In Fig.~\ref{fig:entanglement_vs_T} we illustrate the temperature dependence of $\lambda_E$ for $N = 5$ and $N = 7$ and compare it with the temperature dependence of the classical Lyapunov exponent. Already from Fig.~\ref{fig:ee_vs_time} one can see that the entanglement saturation time is practically independent of the number of degrees of freedom $N_{dof}$ in subsystem $A$. For this reason, Fig.~\ref{fig:entanglement_vs_T} only shows the entanglement saturation time for a single $D0$-brane in order not to clutter the plot. On the other hand, interpreting the entanglement saturation time as a scrambling time in the sense of \cite{Maldacena:1503.01409}, one could expect a mild logarithmic growth $\tau_E \sim \ln\lr{N_{dof}}$ of $\tau_E$ with $N_{dof}$. This observation could either mean that our system is too small for the scaling to be observed, or that entanglement saturation time cannot be identified with the scrambling time. The latter interpretation would make sense at least near the classical limit, because the growth of the coarse-grained entropy is characterized by the Kolmogorov-Sinai entropy, which is roughly proportional to the system size.

We also find that in the high-temperature regime of both for the bosonic matrix model and the full BFSS model the entanglement saturation time $\tau_E \equiv \lambda_E^{-1}$ is very close to the classical Lyapunov time $\tau_L^0 \equiv \lambda_L^0$. This behavior is in sharp contrast to our findings for the quantum Lyapunov exponents, which were extracted from the exponential growth of Lyapunov distances at relatively late times $t \gtrsim \tau_L^0$. On the other hand, entanglement saturation time probes the early-time dynamics at $t \lesssim \tau_L^0$, where Lyapunov distances for the quantum system follow rather closely their classical counterparts. These results suggest that early-time thermalization at high temperatures takes place before the quantum effect becomes important, so that the classical treatment is justified.

As we approach the low-temperature regime the entanglement saturation time for the bosonic matrix model quickly decreases and becomes significantly smaller than the classical Lyapunov exponent, in agreement with the non-chaotic nature of the low-temperature regime. On the other hand, for the full BFSS model the temperature dependence of $\lambda_E$ is far less trivial. In particular, around $T \sim 1$ it exhibits a rather pronounced growth and becomes almost an order of magnitude larger than the classical Lyapunov exponent, thus indicating an extremely fast generation of entanglement. This behavior is most likely an artifact of our approximation, as the BFSS model is not expected to undergo any finite-temperature phase transition \cite{Hanada:0707.4454,Hanada:1802.02985}. In particular, in the full supersymmetric BFSS model fermions prevent the onset of confinement \cite{Hanada:0707.4454,Hanada:1802.02985} which is observed in the purely bosonic model exactly around $T \sim 1$ \cite{Nishimura:0706.3517}. Fast entanglement generation which we observe in our simulation of the BFSS model might be probably interpreted as a kind of ``silver blaze'' phenomenon. Namely, for the full quantum dynamics of the BFSS model fermions conspire to completely remove the signatures of confinement-deconfinement transition from physical observables. However, since in our approximation supersymmetry is not conserved, the influence of fermions on the entanglement dynamics is most likely overestimated, which leads to the artificial speed-up of entanglement generation around the to-be deconfinement temperature of the bosonic matrix model. In this respect this speed-up would be consistent with the general expectation that the near-critical regime of quantum systems is more chaotic and exhibits faster scrambling \cite{Dvali:1507.02948,Susskind:0808.2096}. At even lower temperatures, $\lambda_E$ for the BFSS model again approaches the classical Lyapunov exponent, but never decreases below it.

\begin{figure}
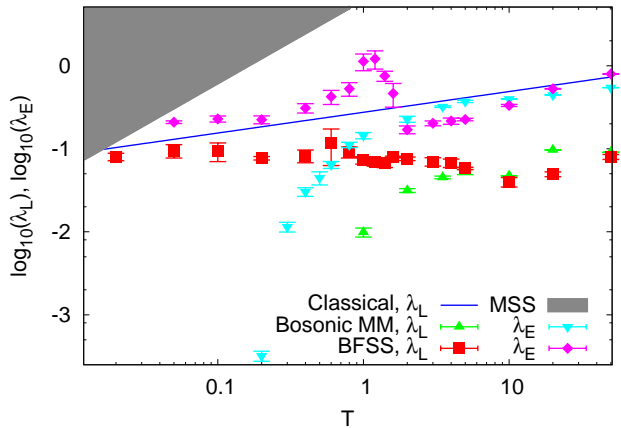

 \centering
 \includegraphics[angle=-90,width=0.48\textwidth]{{{lyapunov_entanglement_summary}}}
 \caption{A comparison of the temperature dependence of the leading Lyapunov exponent $\lambda_L$ and the inverse entanglement saturation time $\lambda_E \equiv \tau_E^{-1}$ in the bosonic matrix model and the BFSS model at $N = 5$. We also sketch the MSS bound $\lambda_L < 2 \pi T$ as well as the classical Lyapunov exponent given by (\ref{classic_bfss_lyapunov}).}
 \label{fig:lyapunov_entanglement_summary}
\end{figure}

Finally, in Fig.~\ref{fig:lyapunov_entanglement_summary} we compare the temperature dependence of the entanglement saturation time $\tau_E \equiv \lambda_E^{-1}$ with that of the characteristic Lyapunov time $\tau_L \equiv \lambda_L^{-1}$, as well as with the MSS bound. The most notable feature is that the entanglement saturation time is always shorter than the characteristic Lyapunov time. These two characteristic timescales become very close to each other and also to the classical Lyapunov time only in the low-temperature regime of the full BFSS model.

\subsection{Quasinormal ringing and quasinormal frequencies}
\label{subsec:qnfs}

While Lyapunov exponents and entanglement generation define the characteristic times for the onset of chaos and spreading of quantum information, the diffusion-driven approach to thermal equilibrium is characterized by another timescale $\tau_D$ set by the decay rate of the quasinormal ringing and thus related to the imaginary parts of quasinormal frequencies.

\begin{figure*}
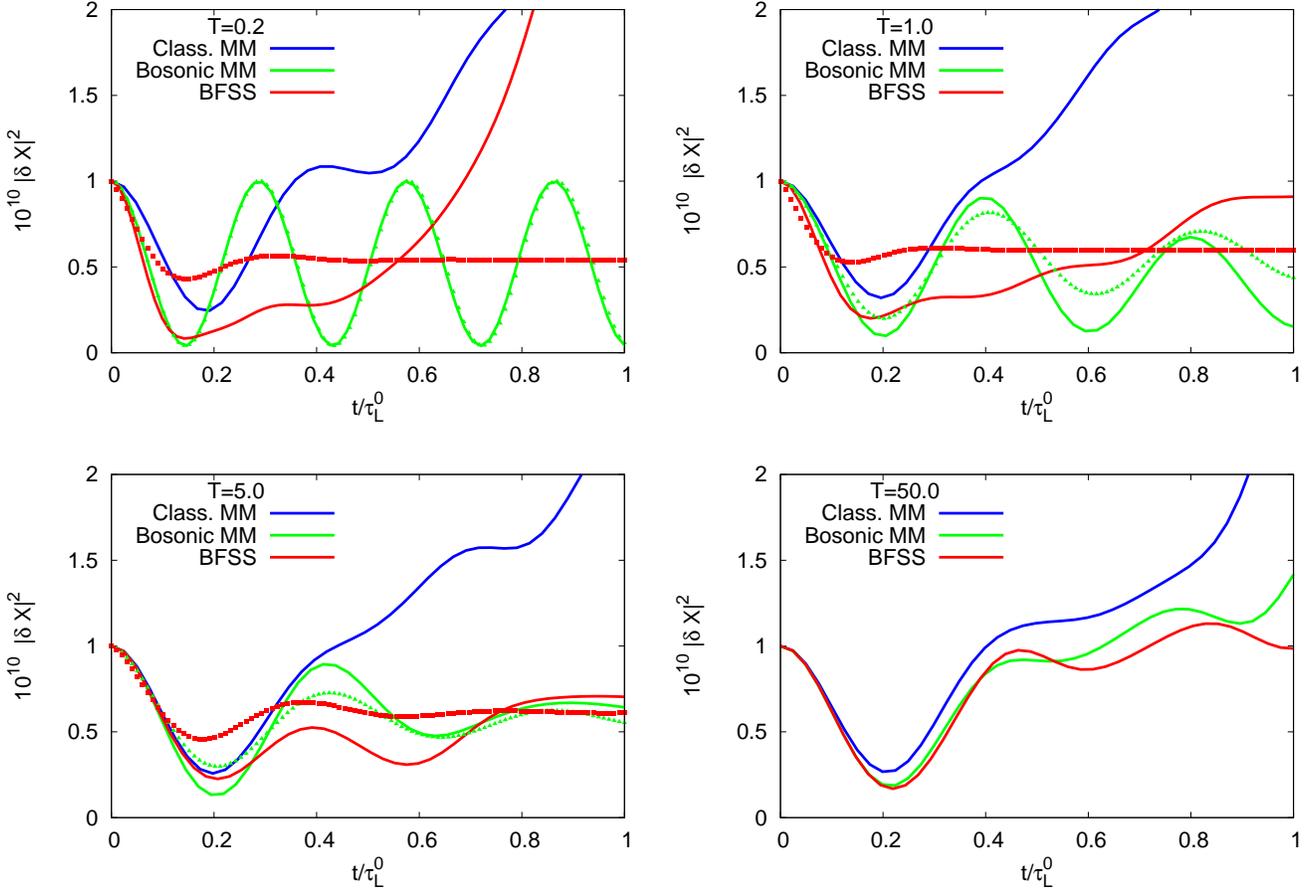

 \centering
 \includegraphics[angle=-90,width=0.49\textwidth]{{{lyapunov_dist_T0.2_enlarged}}}
 \includegraphics[angle=-90,width=0.49\textwidth]{{{lyapunov_dist_T1.0_enlarged}}}\\
 \includegraphics[angle=-90,width=0.49\textwidth]{{{lyapunov_dist_T5.0_enlarged}}}
 \includegraphics[angle=-90,width=0.49\textwidth]{{{lyapunov_dist_T50.0_enlarged}}}\\
 \caption{Signatures of ``quasinormal ringing'' in the early-time evolution of the Lyapunov distances $|\delta X^a_i|^2$ for the classical matrix mechanics, bosonic matrix model and the full BFSS model with $N=5$ at different temperatures. We plot the results for a single instance of random initial conditions and a single random initial Lyapunov shift $\epsilon^a_i$ in (\ref{our_lyapunov_def}). Where the fits of the form (\ref{qnf_ideal_time}) make sense, we show these fits with dotted lines.}
 \label{fig:lyapunov_dist_enlarged}
\end{figure*}

In our numerical study of Lyapunov distances we drive the system out of dynamical equilibrium state by introducing a small coordinate shift $\epsilon^a_i$, see equation (\ref{our_lyapunov_def}). When the ringing decays, the system again reaches dynamical equilibrium, but with a slightly different state within the micro-canonical ensemble than the initial one. At later times, these states become exponentially far separated, and the exponential growth of the Lyapunov distance sets in. Since the timescale of quasinormal ringing appears to be shorter than the Lyapunov time, we can observe a rather clear decay of quasinormal ringing in the time dependence of Lyapunov distances, as one can see already from Fig.~\ref{fig:lyapunov_dist}. On Fig.~\ref{fig:lyapunov_dist_enlarged} we present a better illustration of the quasinormal ringing by showing the time dependence of Lyapunov distances at early times $t < \tau_L^0$ only. The ringing behavior is most clear for the bosonic matrix model. For the full BFSS model and especially for the classical dynamics the onset of exponential growth is sometimes so fast that the ringing cannot be clearly identified.

Nevertheless, it is interesting to study the temperature behavior of quasinormal frequencies at least at the qualitative level. To this end we first identify the duration of the ringing as the period of time during which each successive maximum of the Lyapunov distance $|\delta X^a_i|^2\lr{t}$ is smaller than the previous one. The mean distance $\overline{\Delta t} = \overline{t^{\star}_{i+1} - t^{\star}_i}$ between successive extrema $t^{\star}_i$ of $|\delta X^a_i|^2\lr{t}$ within this time interval is then used to estimate the real part of the quasinormal frequency as $\Re\lr{w} = \pi/\overline{\Delta t}$. We use a simple arithmetic mean over all the extrema which lie within the time interval where the ringing can be identified. To obtain the imaginary part of the quasinormal frequency, we consider the differences $\Delta y_i = |\delta X^a_i|^2\lr{t^{\star}_{i+1}} - |\delta X^a_i|^2\lr{t^{\star}_{i}}$, and estimate
\begin{eqnarray}
\label{im_qnf_estimate}
 \Im\lr{w} = \overline{\ln\lr{\frac{\Delta y_i}{\Delta y_{i+1}}}}/\overline{\Delta t} ,
 \quad
 i = 1, 3, 5, \ldots
\end{eqnarray}
where the mean is again an arithmetic mean over all successive extrema for which $\Delta y_i$ shows the characteristic decaying behavior. This prescription for estimating quasinormal frequencies yields exact results for the quasinormal ringing of the form
\begin{eqnarray}
\label{qnf_ideal_time}
 \vev{\hat{\mathcal{O}}} = \vev{\hat{\mathcal{O}}}_0 + A e^{-\Im w \, t} \cos\lr{\Re w \, t} ,
\end{eqnarray}
where $\hat{\mathcal{O}}$ is some physical observable and $\vev{\hat{\mathcal{O}}}_0$ is its thermal expectation value. On Fig.~\ref{fig:lyapunov_dist_enlarged} we also show the fits of the form (\ref{qnf_ideal_time}) as dotted lines, for those data sets for which such fits can be obtained using the above prescription. In practice this means that we can identify a sufficiently large number of consecutively decaying extrema. For a reliable identification of quasinormal frequencies with smallest imaginary parts (which thus correspond to quasinormal modes with the longest decay time) one needs to analyze the quasinormal ringing at sufficiently late times. For our simulation setup such late times cannot be reached for most data sets due to onset of exponential Lyapunov growth, thus our estimates (\ref{im_qnf_estimate}) of $\Im\lr{w}$ are most likely biased towards somewhat larger values.

\begin{figure*}
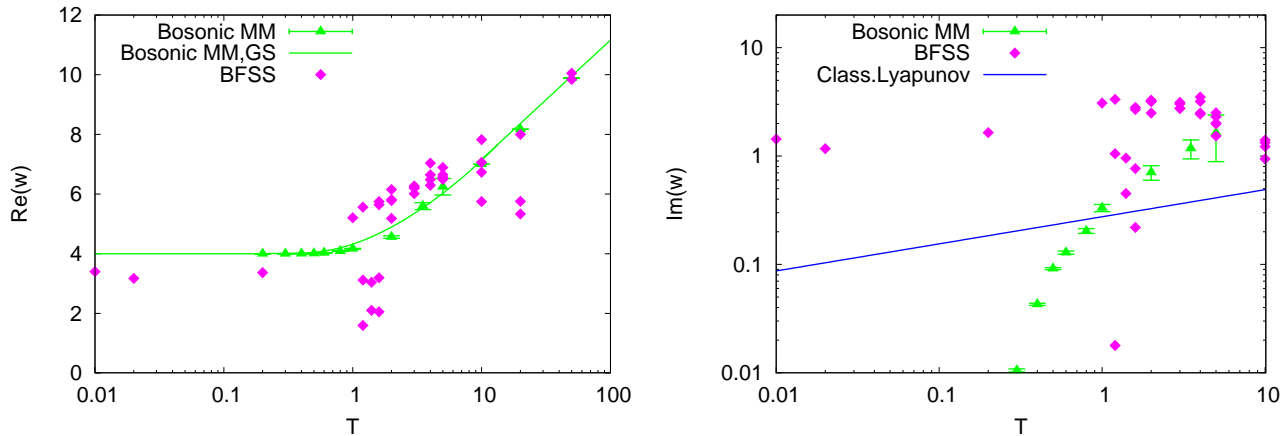

 \centering
 \includegraphics[angle=-90,width=0.48\textwidth]{{{qnf_Re_vs_T}}}
 \includegraphics[angle=-90,width=0.48\textwidth]{{{qnf_Im_vs_T}}}\\
 \caption{Temperature dependence of the real (on the left) and imaginary (on the right) parts of the quasinormal frequency $w_X$ as obtained from the early-time behavior of the Lyapunov distances for the bosonic matrix model and the full BFSS model with $N = 5$. For comparison we also show the temperature dependence of the classical Lyapunov exponent $\lambda_L^0$ given by (\ref{classic_bfss_lyapunov}).}
 \label{fig:qnfs_vs_T}
\end{figure*}

We show our estimates for the real and imaginary parts of the quasinormal frequency $w_X$ associated with the decay of perturbations of $X^a_i$ in the bosonic matrix model and in the BFSS model with $N=5$ on Fig.~\ref{fig:qnfs_vs_T}. For the bosonic matrix model we show the mean values of quasinormal frequencies and the corresponding statistical errors, which can be estimated due to relatively good quality of fits. For the full BFSS model the fits are less reliable, and the standard statistical error does not reflect the fitting uncertainty well. For this case, we thus present only scatter plots of our estimates of $\Re\lr{w}$ and $\Im\lr{w}$ for different random initial conditions.

The temperature dependence of the real part of $w_X$ for the bosonic matrix model is very well approximated by our estimate (\ref{frequency_smallX}) obtained by linearizing equations (\ref{gs_onepoint_eqs}) in the vicinity of thermal state (solid line in the left plot in Fig.~\ref{fig:qnfs_vs_T}). To be more precise, due to the squaring of $\delta X^a_i$ in our definition of Lyapunov distance we observe oscillations at frequency which is exactly two times larger than in (\ref{frequency_smallX}), thus in Fig.~\ref{fig:qnfs_vs_T} we plot $2 w_X$.

For the full BFSS model the temperature dependence of $w_X$ cannot be estimated analytically, as linearization of equations (\ref{gs_onepoint_eqs}) and $\ref{gs_twopoint_eqs}$ in the vicinity of the maximally symmetric solution (\ref{bfss_exact_initial}) with $X^a_i = 0$ and $P^a_i = 0$ would yield exactly the same result as for the bosonic matrix model. To see the difference we should switch to the interpretation of thermal states in terms of dynamical equilibrium states, where the dynamics cannot be treated analytically. Nevertheless, from the left plot in Fig.~\ref{fig:qnfs_vs_T} we see that the temperature dependence of $\Re\lr{w_X}$ for the BFSS model is quite similar to the one for the bosonic matrix model.

The imaginary part of $w_X$ at sufficiently high temperatures turns out to be noticeably larger than the classical Lyapunov exponent. For the bosonic matrix model $\Im\lr{w_X}$ quickly approaches zero at low temperatures, following basically the same trend as for the Lyapunov exponents and the entanglement generation rate. This behavior is in agreement with the general expectation that confining gauge theories are less dissipative, as exemplified, for instance, by the temperature dependence of electric conductivity in QCD \cite{Aarts:1412.6411}.

In contrast, for the full BFSS model imaginary parts of quasinormal frequencies are much larger and seem to remain finite even at the lowest temperatures which we consider, which is again in agreement with the absence of confinement regime in the BFSS model. While at very high temperatures we cannot reliably extract $w_X$ from the data, the plot for $T = 50.0$ on Fig.~\ref{fig:lyapunov_dist_enlarged} makes it clear that at this temperature the quasinormal ringing is very similar for the classical matrix mechanics, the bosonic matrix model and the BFSS model, in agreement with the classicality of the high-temperature limit. The plots on Fig.~\ref{fig:lyapunov_dist_enlarged} also suggest that $\Im\lr{w_X}$ for the classical theory is still larger than for the BFSS model, except probably for the lowest temperatures. For the bosonic matrix model $\Im\lr{w_X}$ takes smallest values.

Finally, let us note that at high temperatures we can use equations (\ref{frequency_smallX}), (\ref{frequency_smallXX}), (\ref{temperature_def_highT}) and (\ref{equilibrium_vals_vs_f}) to estimate the real part of the quasinormal frequency of the quasinormal ringing of $\tr X_i^2$ as $w_{XX} = 4.89 \, T^{1/4}$, which agrees well with the dependence $w_{XX} = 5.152 \, T^{1/4}$ obtained in \cite{Aprile:1611.00786,HanadaTalkKyoto} for the bosonic matrix model using more elaborate dedicated simulations. At the highest temperatures where $\Im\lr{w_X}$ can still be estimated in our simulations, we obtain $\Im\lr{w_X}/\Re\lr{w_X} \sim 0.1$, which also roughly agrees with the result of \cite{Aprile:1611.00786,HanadaTalkKyoto}. According to the dual gravity calculation, near zero temperature the quasinormal frequency is proportional to $T$ and the ratio $\Im\lr{w_X}/\Re\lr{w_X}$ is about 1.7 \cite{Iizuka:hep-th/0306209}. While our calculation is too crude to capture the result near $T=0$, the observed growth of this ratio follows the right trend.

\section{Conclusions and outlook}
\label{sec:conclusions}

Until recently the real-time dynamics of the BFSS model could only be addressed either in the low-energy regime, which is tractable in terms of the dual holographic description, or in the high-energy regime in which the system becomes effectively classical. Our simulations of the corresponding ungauged models \cite{Maldacena:1802.00428,Hanada:1802.02985} within the Gaussian state approximation is a step towards bridging the gap between these two regimes.

The Gaussian state approximation appears to be much more accurate for the bosonic matrix model than for the BFSS model, as suggested by the quantitatively good agreement of our equation of state (\ref{equilibrium_vals_vs_f}) with the results of first-principle Monte-Carlo simulations \cite{Hanada:1802.02985}. Our results for the bosonic model are thus likely to be reliable at the quantitative level, while for the BFSS model we can make at most qualitative statements.

We have explicitly studied and confirmed some of the important features of the real-time dynamics of the bosonic matrix model and the BFSS model which fit expectations either based on general grounds \cite{Susskind:0808.2096} or motivated by the dual holographic description \cite{Witten:hep-th/9510135,Nicolai:NPB1988}:
\begin{itemize}
 \item Quantum real-time dynamics is characterized by smaller Lyapunov exponents in comparison with the classical system at the same energy. This ensures the validity of the MSS bound \cite{Maldacena:1503.01409} for all energies.
 \item Gauged bosonic matrix model becomes confining and non-chaotic at low temperatures \cite{Nishimura:0706.3517,Aharony:hep-th/0406210}, with Lyapunov exponents being $1/N$-suppressed.
 At low temperature the gauged and ungauged model should become exponentially close \cite{Maldacena:1802.00428,Hanada:1802.02985}, and hence the Lyapunov exponent of the ungauged model should be exponentially small. We indeed observe such a suppression of Lyapunov exponents in our simulations. Decay time of quasinormal modes and the saturation time for entanglement entropy also become very long.
 \item When fermionic degrees of freedom are added, the BFSS model exhibits chaotic behavior and fast decay of quasinormal modes at all temperatures, in agreement with the absence of confining regime all the way down to zero temperature \cite{Hanada:0707.4454,Hanada:1802.02985}. The Lyapunov exponents, however, still appear to be smaller than in the classical theory at all temperatures.
 \item Entanglement entropy shows the expected ``scrambling'' behavior \cite{Susskind:0808.2096}: the initial growth followed by saturation at later times. For sufficiently small subsystems, late-time entanglement entropy per degree of freedom for a single pure state in the micro-canonical ensemble is equal to the von Neumann entropy per degree of freedom for the thermal density matrix, which illustrates an equivalence between micro-canonical and canonical ensemble for quantum-chaotic system \cite{Susskind:0808.2096}.
 \item The characteristic time $\tau_E$ at which the entanglement entropy reaches saturation is shorter than the quantum Lyapunov time $\tau_L \equiv \lambda_L^{-1}$ (see the discussion below for a possible loophole), and also much closer to the classical Lyapunov time at high temperatures. In the intermediate- and low-temperature regime of the BFSS model, $\tau_E$ appears to be close to or even shorter than the classical Lyapunov time, in agreement with the chaoticity of the BFSS model at all temperatures. Nevertheless, $\tau_E$ is still longer than the decay time $\tau_D$ of quasinormal modes.
\end{itemize}
Figs.~\ref{fig:lyapunov_entanglement_summary} and \ref{fig:qnfs_vs_T} provide a compact illustration of these findings.

Our results for the time dependence of the entanglement entropy are under best theoretical control, as they are extracted from the early-time behavior for which the Gaussian state approximation should be quantitatively accurate. At the same time, the relatively smooth time dependence allows for rather unambiguous definition of the entanglement saturation time $\tau_E$. Quasinormal frequencies are also extracted from early-time behavior, however, the extraction of the decay rate of quasinormal modes is not so reliable and unambiguous. On the other hand, Lyapunov distances are extracted from the behavior of Lyapunov distances at relatively late times, where the Gaussian state approximation can be accurate at most qualitatively. Thus while the exponential fits to the time dependence of Lyapunov distances are quite good, our estimates of Lyapunov exponents can still be biased, especially at high temperatures.

Our real-time analysis might be further improved if one considers correlators with more than two canonical variables, for example, within the $n$-particle irreducible effective action techniques \cite{Berges:hep-ph/0401172}. While for higher-dimensional gauge theories the application of such methods faces the difficulty of parameterizing and storing higher-order correlators which depend on many momenta, low dimensionality and numerous (super)symmetries of the BFSS model (\ref{bfss_Hamiltonian}) should significantly reduce the number of independent correlation functions which enter the analysis.

In fact, we have already tried to extend equations (\ref{gs_onepoint_eqs}) and (\ref{gs_twopoint_eqs}) by incorporating the connected correlator $\cev{\hpsi^a_{\alpha} \hpsi^b_{\beta} \hX^c_i}$ and requiring the solution to be $SU\lr{N}$ and rotationally symmetric. Unfortunately, it turned out that such an extension is not intrinsically consistent in that the correlators $\cev{\hX^a_i \hX^b_j}$ cease to be positive-definite after a relatively short evolution time. This indicates the necessity to include correlators of even higher orders into the analysis.

It would be also interesting to understand the applicability of the Gaussian state approximation to real-time dynamics of higher-dimensional gauge theories, as well as its interpretation in the context of the conventional scale separation between soft-momenta and hard-momenta gauge fields which allows to treat the dynamics of soft gauge fields classically. In principle, Gaussian state approximation should extend the range of validity of real-time simulations of Yang-Mills theory beyond that of the classical dynamics at a numerical cost which scales quadratically with spatial volume, which is thus comparable with the numerical cost of real-time simulations with fermions \cite{Borsanyi:0809.4711,Berges:1403.4849}. In particular, in contrast to purely classical dynamics the Gaussian state approximation should be able to incorporate transitions between different topological sectors, and might thus provide an alternative to real-time evolution equations which include Langevin noise to account for the contribution of hard gauge fields \cite{Bodeker:hep-ph/9801430,Son:hep-ph/9810216}, in particular for problems like the estimation of sphaleron rate \cite{Rothkopf:1512.02374}.

\begin{acknowledgments}
P.~B. is supported by the Heisenberg Fellowship from the German Research Foundation, project BU2626/3-1. M.~H. acknowledges JSPS KAKENHI Grants 17K14285. This work was also partially supported by the Department of Energy, award number DE-SC0017905. The calculations were performed on the ``iDataCool'' cluster at Regensburg University and on the LRZ cluster in Garching. We acknowledge valuable discussions with D.~Berenstein, N.~Bodendorfer, D.~{O'Connor}, P.~Romatschke and A.~Rothkopf.
\end{acknowledgments}


%

\appendix

\section{Conserved quantities and supersymmetry for BFSS Hamiltonian in the Gaussian state approximation}
\label{apdx:csft_symmetries}

Equations of motion (\ref{gs_onepoint_eqs}) and (\ref{gs_twopoint_eqs}) conserve energy and gauge constraint, which are given by the expectation values of the operators (\ref{bfss_Hamiltonian}) and (\ref{gauge_generators}). Yet another obvious integral of motion for the Hamiltonian (\ref{bfss_Hamiltonian}) is the total angular momentum, described by the operator
\begin{eqnarray}
\label{angular_momentum}
 \hat{J}_{ij} = \hX^a_i \hP^a_j - \hX^a_j \hP^a_i
 -
 \frac{i}{8} \hpsi^a_{\alpha} \, \sigma_{ij}^{\alpha\beta} \, \hpsi^a_{\beta} ,
\end{eqnarray}
which acts on the vectors and spinors as $\lrs{\hat{J}_{kl}, \hX^a_i} = i \delta_{ik} \hX^a_l - i \delta_{il} \hX^a_k$ and $\lrs{\hat{J}_{kl}, \hpsi^a} = \frac{i}{4} \sigma_{kl} \hpsi^a$.

Here we give explicit expressions for these conserved quantities:
\begin{eqnarray}
\label{gs_conserved_energy}
E
\equiv
\vev{\hat{H}}
 = \nonumber\\ =
 \frac{P^a_i P^a_i + \cev{\hP^a_i \hP^a_i}}{2 N}
 + \nonumber \\ +
 \frac{N}{4} C_{abc} C_{ade}
 \left(
 X^b_i X^c_j X^d_i X^e_j
 + \right. \nonumber \\ +
 \cev{\hX^b_i \hX^c_j} \cev{\hX^d_i \hX^e_j}
 + \nonumber \\ +
 \cev{\hX^b_i \hX^d_i} \cev{\hX^c_j \hX^e_j}
 + \nonumber \\ +
 \cev{\hX^b_i \hX^e_j} \cev{\hX^d_i \hX^c_j}
 +
 \cev{\hX^b_i \hX^c_j} X^d_i X^e_j
 + \nonumber \\ +
 \cev{\hX^b_i \hX^d_i} X^c_j X^e_j
 +
 \cev{\hX^b_i \hX^e_j} X^d_i X^c_j
 + \nonumber \\ +
 \cev{\hX^d_i \hX^e_j} X^b_i X^c_j
 +
 \cev{\hX^c_j \hX^e_j} X^b_i X^d_i
 + \nonumber \\ \left. +
 \cev{\hX^d_i \hX^c_j} X^b_i X^e_j
 \right) ,
\end{eqnarray}
\begin{eqnarray}
\label{gs_gauge_generators}
 J^a \equiv
 \vev{\hat{J}^a}
 = \nonumber\\ =
 C_{abc} X^b_i P^c_i
 +
 C_{abc} \cev{ \hX^b_i \hP^c_i }
 - \nonumber \\ -
 \frac{i}{2} C_{abc} \cev{ \hpsi^b_{\alpha} \hpsi^c_{\alpha} } ,
\end{eqnarray}
\begin{eqnarray}
\label{gs_angular_momentum}
 J_{ij} \equiv 
 \vev{ \hat{J}_{ij} }
 = \nonumber\\ =
 X^a_i P^a_j - X^a_j P^a_i
 +
 \cev{ \hX^a_i \hP^a_j }
 - \nonumber\\ - 
 \cev{ \hX^a_j \hP^a_i }
 -
 \frac{i}{8} \, \sigma_{ij}^{\alpha\beta} \, \cev{\hpsi^a_{\alpha} \hpsi^a_{\beta}} .
\end{eqnarray}
An explicit demonstration of the conservation of these quantities from equations (\ref{gs_onepoint_eqs}) and (\ref{gs_twopoint_eqs}) is a lengthy but straightforward calculation, which we do not present here in order to save space.

The conservation of the supersymmetry generators $\hat{Q}_{\alpha}$ in the Gaussian state approximation is a more subtle question. On the one hand, the expectation values $\vev{\hat{Q}_{\alpha}}$ are zero for our Gaussian state with $\vev{\hpsi} = 0$ and vanishing mixed bosonic-fermionic correlators, and are hence trivially conserved. On the other hand, if the Gaussian state approximation preserved supersymmetry, we should have obtained a zero ground state energy. Since this is not the case, supersymmetry should be somehow violated.

To understand the fate of supersymmetry in more details, let us first present an outline of the proof of the conservation of supersymmetry generators (\ref{bfss_susy_charge}) from the full Heisenberg equations (\ref{heisenberg_eqs}), which after some algebra yield
\begin{eqnarray}
\label{susy_charge_full_evolution}
 \partial_t \hat{Q}_{\delta}
 =
 C_{abc} P^b_i X^c_i \hpsi^a
 -
 \frac{i}{2} C_{bac} \hpsi^a_{\gamma} \hpsi^b_{\alpha} \hpsi^c_{\beta}
 \sigma^i_{\alpha\beta} \sigma^i_{\gamma\delta} .
\end{eqnarray}
Taking into account the vanishing of the gauge constraint (\ref{gauge_generators}) on the physical Hilbert space, we can also rewrite (\ref{susy_charge_full_evolution}) as
\begin{eqnarray}
\label{susy_charge_full_evolution1}
 \partial_t \hat{Q}_{\delta} = \frac{i}{2} C_{abc} \hpsi^a_{\alpha} \hpsi^b_{\beta} \hpsi^c_{\gamma} \lr{\sigma^i_{\alpha\beta} \sigma^i_{\gamma\delta} - \delta_{\alpha\beta} \delta_{\gamma\delta} } .
\end{eqnarray}
Now one can use the anti-commutativity of the operators $\hpsi^a_{\alpha}$ and the cyclic symmetry of the structure constants $C_{abc} = C_{bca} = C_{cab}$ to transform this expression into
\begin{eqnarray}
\label{susy_charge_full_evolution2}
 \partial_t \hat{Q}_{\delta}
 &=&
 \frac{i}{6} C_{abc} \hpsi^a_{\alpha} \hpsi^b_{\beta} \hpsi^c_{\gamma}
  \nonumber \\
  & &\times
 \left(
  \sigma^i_{\alpha\beta} \sigma^i_{\gamma\delta}
  +
  \sigma^i_{\beta\gamma} \sigma^i_{\alpha\delta}
  +
  \sigma^i_{\gamma\alpha} \sigma^i_{\beta\delta}
  \right. \nonumber \\
  & &
  \quad\left. -
  \delta_{\alpha\beta} \delta_{\gamma\delta}
  -
  \delta_{\beta\gamma} \delta_{\alpha\delta}
  -
  \delta_{\gamma\alpha} \delta_{\beta\delta}
 \right) .
\end{eqnarray}
In other words, the indices $\alpha$, $\beta$ and $\gamma$ are cyclically permuted. But precisely this combination of $\sigma$ matrices is equal to zero by virtue of the Fierz identity (\ref{fierz_identity}). Hence the time derivative $\partial_t \hat{Q}_{\alpha}$ is equal to zero on the physical Hilbert space.

Now we turn to the Gaussian state approximation and take the limit in which the classical expectation values $X^a_i$, $P^a_i$ are so large that their quantum dispersions can be neglected. In this case the evolution of the system is described by the classical equations of motion for $X^a_i$ and $P^a_i$, augmented by the fermionic force, and fermions evolve in the classical background of $X^a_i$ variables. This corresponds to the classical-statistical field theory approximation (CSFT), which is a standard method for addressing the real-time dynamics of fermions interacting with strong gauge fields, see e.g.~\cite{Berges:1403.4849}. Since the fermionic part of the BFSS Hamiltonian (\ref{bfss_Hamiltonian}) is quadratic in the fermionic fields, the fermionic wave function is always Gaussian, and is in one-to-one correspondence with the fermionic correlators $\cev{\hpsi^a_{\alpha} \hpsi^b_{\beta}}$. In this case we can simplify calculations by working directly with the operators $\hpsi^a_{\alpha}$ which satisfy the Heisenberg equations of motion. In this limit the supersymmetry generators (\ref{bfss_susy_charge}) take the form
\begin{eqnarray}
\label{csft_susy_charge}
 \hat{Q}_{\alpha} = P^a_i \sigma^i_{\alpha\beta} \hpsi^a_{\beta} - \frac{N}{4} C_{a d e} X^d_i X^e_j \sigma^{ij}_{\alpha\beta} \hpsi^a_{\beta} .
\end{eqnarray}
Using equations of motion (\ref{gs_onepoint_eqs}), after some algebra we can represent the time derivative of $\hat{Q}_{\alpha}$ as
\begin{eqnarray}
\label{csft_susy_charge_evolution}
 \partial_t \hat{Q}_{\delta} = \frac{i}{2} C_{abc} \cev{ \hpsi^a_{\alpha} \hpsi^b_{\beta} } \hpsi^c_{\gamma} \lr{\sigma^i_{\alpha\beta} \sigma^i_{\gamma\delta} - \delta_{\alpha\beta} \delta_{\gamma\delta} } .
\end{eqnarray}
The only difference with the expression (\ref{susy_charge_full_evolution}) for the full quantum evolution is that two out of three $\hpsi$ operators are now under the vacuum expectation value brackets. This vacuum expectation value is in fact the expectation value of the fermionic force term $\frac{i}{2} C_{b a c} \sigma^i_{\alpha\beta} \cev{\hpsi^b_{\alpha} \hpsi^c_{\beta}}$ on the right-hand side of equation (\ref{gs_onepoint_eq_P}) which governs the time evolution of $P^a_i$.

It is now easy to see that due to these brackets, one can no longer cyclically permute different $\hpsi$ operators and use the Fierz identities (\ref{fierz_identity}) as in (\ref{susy_charge_full_evolution2}) to show the conservation of the SUSY charge. One can construct a similar proof by considering arbitrary Gaussian states with $\vev{\hpsi^a_{\alpha}} \neq 0$ and $\vev{\hpsi^a_{\alpha} \hX^{b}_i} \neq 0$, $\vev{\hpsi^a_{\alpha} \hP^{b}_i} \neq 0$ which mix fermionic and bosonic variables. This makes the derivation significantly more involved, but leads to similar conclusions. We thus conclude that supersymmetry is not preserved in the Gaussian state approximation.

\section{Ground state and real-time evolution of Majorana fermions}
\label{apdx:majorana_initial_state}

In our approximation the thermodynamics and the real-time evolution of Majorana fermions $\hpsi^a_{\alpha}$ are governed by the fermionic Hamiltonian
\begin{eqnarray}
\label{fermionic_hamiltonian}
 \hat{H}_F
 &=&
 \frac{1}{2} h^{a b}_{\alpha\beta} \hpsi^a_{\alpha} \hpsi^b_{\beta} ,
 \\
\label{fermionic_sp_hamiltonian}
 h_{a b}^{\alpha\beta}
 &=&
 i C_{a c b} X^c_i \sigma_i^{\alpha\beta}  ,
\end{eqnarray}
where we have introduced a single-particle Hamiltonian $h^{a b}_{\alpha\beta}$ which defines the energy levels occupied by Majorana fermions.

In order to describe the ground state of $\hat{H}_F$, we note that the single-particle Hamiltonian (\ref{fermionic_sp_hamiltonian}) is a real anti-symmetric matrix of size $N_F = 16 \lr{N^2 - 1}$ multiplied by the unit imaginary number $i$. The eigenvalues of real anti-symmetric matrices with even size are purely imaginary and come in conjugate pairs $\pm i \lambda_k$, $k = 1 \ldots N_F/2$. Correspondingly, the energy levels of the single-particle Hamiltonian (\ref{fermionic_sp_hamiltonian}) come in opposite-sign pairs $+\epsilon_k$, $-\epsilon_k$, which reflects the particle-anti-particle symmetry of Majorana fermions. In what follows we assume that the index $k = 1 \ldots N_F/2$ enumerates the positive energy levels $\epsilon_k > 0$.

It is easy to see that the eigenvectors which correspond to energies $+ \epsilon_k$ and $-\epsilon_k$ are complex conjugate to each other. Splitting these eigenvectors into real and imaginary parts $\psi\lr{+\epsilon_k} = u_k\lr{\epsilon_k} + i v_k\lr{\epsilon_k}$, $\psi\lr{-\epsilon_k} = u_k\lr{\epsilon_k} - i v_k\lr{\epsilon_k}$, $k = 1 \ldots N_F/2$, we write the eigenvalue equation for the single-particle Hamiltonian (\ref{fermionic_sp_hamiltonian}) as
\begin{eqnarray}
\label{eigenbasis_def}
 h_{ab}^{\alpha\beta} u^b_{\beta}\lr{\epsilon_k}
 &=&
 i \epsilon_k \, v^a_{\alpha}\lr{\epsilon_k},
 \nonumber \\
 h_{ab}^{\alpha\beta} v^b_{\beta}\lr{\epsilon_k}
 &=&
 - i \epsilon_k \, u^a_{\alpha}\lr{\epsilon_k} ,
\end{eqnarray}
Completeness and orthogonality of the basis of complex-valued eigenvectors $\psi\lr{\pm \epsilon_k}$ also imply the following completeness and orthogonality relations for $u$ and $v$:
\begin{eqnarray}
\label{uv_orthogonality}
& &
u\lr{\epsilon_k} \cdot u\lr{\epsilon_l} + v\lr{\epsilon_k} \cdot v\lr{\epsilon_l} = \delta_{kl} ,
 \nonumber \\
& &
u\lr{\epsilon_k} \cdot v\lr{\epsilon_l} - v\lr{\epsilon_k} \cdot u\lr{\epsilon_l} = 0 ,
 \\
\label{uv_completeness}
& & \sum\limits_k \lr{u\lr{\epsilon_k} \otimes u\lr{\epsilon_k} + v\lr{\epsilon_k} \otimes v\lr{\epsilon_k}} = I/2 ,
\end{eqnarray}
There are more relations which can be obtained by taking the scalar products of eigenvectors with opposite signs of $\epsilon$, but we will not need them here.

Decomposing the operators $\hpsi^a_{\alpha}$ in the basis of real vectors $u^a_{\alpha}\lr{\epsilon_k}$, $v^a_{\alpha}\lr{\epsilon_k} \hpsi_k^{\lr{2}}$ as
\begin{eqnarray}
\label{psi_ansatz}
& &
\hpsi^a_{\alpha} = \sqrt{2} \, \sum\limits_k \lr{
   u^a_{\alpha}\lr{\epsilon_k} \hpsi_k^{\lr{1}}
   +
   v^a_{\alpha}\lr{\epsilon_k} \hpsi_k^{\lr{2}}
  },
   \nonumber \\
 & &
 \lrc{\hpsi_k^{\lr{A}}, \hpsi_l^{\lr{B}}} = \delta_{AB} \delta_{kl},
   \quad A, B = 1, 2,
\end{eqnarray}
we can bring our Hamiltonian into the form
\begin{eqnarray}
\label{diagonalized_Hamiltonian}
 \hat{H}_F = -2 i \sum\limits_k \epsilon_k \, \hpsi_k^{\lr{1}} \hpsi_k^{\lr{2}} .
\end{eqnarray}
Upon this decomposition, the many-body fermionic Hamiltonian completely splits into a sum of independent Hamiltonians for each positive energy level $\epsilon_k > 0$. For each $k$ the algebra of operators $\hpsi_k^{\lr{1}}$, $\hpsi_k^{\lr{2}}$ is isomorphic to the algebra of $2 \times 2$ Pauli matrices $s_1$, $s_2$, $s_3$:
\begin{eqnarray}
\label{majorana_pauli_isomorphism}
 \hpsi_k^{\lr{1}}
 &\rightarrow&
 s_1/\sqrt{2},
 \nonumber\\
 \hpsi_k^{\lr{2}}
 &\rightarrow&
 s_2/\sqrt{2},
 \nonumber \\
 -2 i \hpsi_k^{\lr{1}} \hpsi_k^{\lr{2}}
 &\rightarrow&
 s_3 = \diag{+1, -1} .
\end{eqnarray}
This isomorphism allows one to immediately find the thermal partition function
\begin{eqnarray}
\label{maj_ham_partfunc}
 \mathcal{Z} = \tr e^{-\beta \hat{H}_F} = \prod\limits_k 2 \cosh\lr{\beta \epsilon_k}
\end{eqnarray}
as well as the fermionic two-point function
\begin{eqnarray}
\label{maj_ham_twopoint_diag}
 \lefteqn{\mathcal{Z}^{-1} \tr\lr{\hpsi_k^{\lr{A}} \hpsi_l^{\lr{B}} e^{-\beta \hat{H}_F}}
 } \nonumber \\
 &=&
 \delta_{kl} \lr{\frac{\delta_{AB}}{2} - \frac{i \, \varepsilon_{A B}}{2} \tanh\lr{\beta \epsilon_k}},
\end{eqnarray}
where $\varepsilon_{AB}$ is the $2 \times 2$ antisymmetric matrix with ${\varepsilon_{12} = 1}$. In this work we use the correlator (\ref{maj_ham_twopoint_diag}) in the zero-temperature limit $\beta \rightarrow \infty$ as the initial condition for $\cev{\hpsi^a_{\alpha} \hpsi^a_{\beta}}$. Using the definition (\ref{psi_ansatz}) of the operators $\hpsi_k^{\lr{1}}$ and $\hpsi_k^{\lr{2}}$ and replacing $\tanh\lr{\beta \epsilon_k} = 1$ at $\beta \rightarrow \infty$, we obtain an explicit expression for $\cev{\hpsi^a_{\alpha} \hpsi^a_{\beta}}$ in the zero-temperature limit:
\begin{eqnarray}
\label{maj_ham_twopoint}
 \cev{\hpsi^a_{\alpha} \hpsi^b_{\beta}}_0
 &=&
 \frac{\delta^{ab} \delta_{\alpha\beta}}{2}
 -
 i \sum\limits_k
 \left\{
  u^a_{\alpha}\lr{\epsilon_k} v^b_{\beta}\lr{\epsilon_k}
  \right.
   \nonumber \\
 & &
 \qquad
 \left.
  -
  v^a_{\alpha}\lr{\epsilon_k} v^b_{\beta}\lr{\epsilon_k}
 \right\}.
\end{eqnarray}
As a quick check of the above expressions, we note that the expectation value of the fermionic Hamiltonian $\vev{ \hat{H}_F } = \frac{1}{2} h_{a b}^{\alpha\beta} \vev{ \hpsi^a_{\alpha} \hpsi^b_{\beta} }$ is negative, which can be expected if only the negative energy levels are filled by fermions. However, for Majorana fermions which are their own anti-particles this statement is somewhat subtle.

Having fixed the initial conditions for the fermionic correlators $\cev{\hpsi^a_{\alpha} \hpsi^b_{\beta}}$, we have to solve the equation (\ref{gs_twopoint_eqs_psipsi}) which governs the time evolution of this correlator. This can be done by promoting the basis vectors $u^a_{\alpha}\lr{\epsilon}$ and $v^a_{\alpha}\lr{\epsilon}$ to time-dependent functions which satisfy the single-particle Schr\"{o}dinger equations
\begin{eqnarray}
\label{csft_eq_mode_functions}
 \partial_t u^a_{\alpha}\lr{\epsilon_k}
 &=&
 C_{a b c} X^b_i \sigma^i_{\alpha\beta} u^c_{\beta}\lr{\epsilon_k} ,
 \nonumber \\
 \partial_t v^a_{\alpha}\lr{\epsilon_k}
 &=&
 C_{a b c} X^b_i \sigma^i_{\alpha\beta} v^c_{\beta}\lr{\epsilon_k} .
\end{eqnarray}
At $t = 0$, $u^a_{\alpha}\lr{\epsilon_k}$ and $v^a_{\alpha}\lr{\epsilon_k}$ are defined by (\ref{eigenbasis_def}). Since the single-particle Hamiltonian (\ref{fermionic_sp_hamiltonian}) is time-dependent, at $t > 0$ $u^a_{\alpha}\lr{\epsilon_k}$ and $v^a_{\alpha}\lr{\epsilon_k}$ are in general no longer related to the eigenstates of $h_{ab}^{\alpha\beta}$. It is easy to check that the two-point function (\ref{maj_ham_twopoint}) with time-dependent functions $u^a_{\alpha}\lr{\epsilon_k}$ and $v^a_{\alpha}\lr{\epsilon_k}$ which satisfy equations (\ref{csft_eq_mode_functions}) solves equation (\ref{gs_twopoint_eqs_psipsi}). In the literature on real-time simulations within the classical-statistical field theory (CSFT) approximation the functions $u^a_{\alpha}\lr{\epsilon_k}$ and $v^a_{\alpha}\lr{\epsilon_k}$ are commonly referred to as mode functions \cite{Borsanyi:0809.4711,Berges:1403.4849}.

This way of solving equation (\ref{gs_twopoint_eqs_psipsi}) requires at least two times less memory and CPU time than a straightforward solution. For illustration, we present the equation of motion for the bosonic momenta $P^a_i$ in terms of mode functions:
\begin{eqnarray}
\label{csft_eq_momentum}
 \partial_t P^a_i
 = - C_{abc} C_{cde} X^b_j X^d_i X^e_j
 + \nonumber \\ +
 \sum\limits_k C_{a b c} \sigma^i_{\alpha\beta} u^b_{\alpha}\lr{\epsilon_k} v^c_{\beta}\lr{\epsilon_k}.
\end{eqnarray}

\section{Symplectic structure in the Gaussian state approximation}
\label{apdx:symplectic_conservation}

As discussed in Subsection~\ref{subsec:entanglement_entropy}, a necessary and sufficient condition for a general Gaussian density matrix to describe a pure state is that the corresponding correlator matrix (\ref{correlator_block_matrix}) should have symplectic eigenvalues all equal to $f_k = 1/2$. In this Appendix we demonstrate that equations (\ref{gs_onepoint_eqs}) and (\ref{gs_twopoint_eqs}) conserve symplectic eigenvalues of the correlator matrix (\ref{correlator_block_matrix}) and hence evolve pure states into pure states. For mixed states this property obviously implies the conservation of von Neumann and R\'{e}nyi entropies.

To begin with, we introduce the condensed index notation $A = \lrc{a, i}$, $B = \lrc{b, j}$ and rewrite equations (\ref{gs_twopoint_eqs}) as
\begin{eqnarray}
\label{csft_quartic_2p_eqs}
 \partial_t \cev{\hX_A \hX_B}
 &=&
 \cev{\hX_A \hP_B} + \cev{\hP_A \hX_B} ,
 \nonumber \\
 \partial_t \cev{\hX_A \hP_B}
 &=&
 \cev{ \hP_A \hP_B} -
 \nonumber \\
 & &
 -
 3 \cev{\hX_A \hX_E} V_{ECDB} \vev{\hX_C \hX_D} ,
 \nonumber \\
 \partial_t \cev{\hP_A \hP_B}
 &=& -
 3 V_{ACDE} \vev{\hX_C \hX_D} \cev{\hX_E \hP_B}
\nonumber \\
& &
-
 3 V_{BCDE} \vev{\hX_C \hX_D} \cev{\hX_E \hP_A} , \quad
\end{eqnarray}
where $V_{ABCD}$ is the short-hand notation for the coefficients of the quartic term in the Hamiltonian (\ref{bfss_Hamiltonian}), which thus takes the form $\hat{H} = \hP_A^2/2 + V_{ABCD} \hX_A \hX_B \hX_C \hX_D/4$.

It is now convenient to introduce the symmetric and real matrix $V_{AB} = 3 V_{ABCD} \vev{\hX_C \hX_D}$ and to treat the correlators $\cev{\hX_A \hX_B}$, $\cev{\hX_A \hP_B}$ and $\cev{\hP_A \hP_B}$ as matrices, omitting their indices. This allows to write equations (\ref{csft_quartic_2p_eqs}) in a particularly simple form:
\begin{eqnarray}
\label{csft_quartic_eqs_matrix1}
 \partial_t \cev{\hX \hX}
 &=&
 \cev{\hX \hP} + \cev{\hX \hP}^T ,
 \nonumber \\
 \partial_t \cev{\hX \hP}
 &=&
 \cev{\hP \hP} - \cev{\hX \hX} V ,
 \nonumber \\
 \partial_t \cev{\hP \hP}
 &=&
 - V \cev{\hX \hP} - \cev{\hX \hP}^T V .
\end{eqnarray}
We now combine all the correlators in the block matrix $\Delta$ given by (\ref{correlator_block_matrix}), and introduce the block matrix
\begin{eqnarray}
\label{csft_quartic_upsilon_def}
 \Upsilon
 =
 \left(
   \begin{array}{cc}
      0 & I \\
      -V & 0 \\
   \end{array}
 \right) .
\end{eqnarray}
In terms of the symplectic form $\Omega$ defined in (\ref{symplectic_form_def}) and the matrices $\Delta$ and $\Upsilon$, equations (\ref{csft_quartic_eqs_matrix1}) can be written as
\begin{eqnarray}
\label{csft_quartic_eqs_matrix}
 \partial_t \lr{\Delta \Omega} = \Upsilon \lr{\Delta \Omega} - \lr{\Delta \Omega} \Upsilon.
\end{eqnarray}
The commutator structure on the right-hand side implies that the eigenvalues of the matrix $\Delta \Omega$ are conserved during the evolution described by equations (\ref{csft_quartic_2p_eqs}), which are a compact representation of equations (\ref{gs_twopoint_eqs}). Since the eigenvalues of $\Delta \Omega$ are the symplectic eigenvalues of $\Delta$ which determine whether the state is pure or not, we have thus proven that pure states are evolved into pure states. While the form of equations (\ref{csft_quartic_2p_eqs}) is only valid for models with quartic interactions, our proof can be easily generalized for other Hamiltonians. In particular, the effect of Majorana fermions can be easily incorporated as a time-dependent linear potential in addition to the quartic potential.

\section{Numerical discretization of equations (\ref{gs_onepoint_eqs}) and (\ref{gs_twopoint_eqs})}
\label{apdx:discretization}

In order to solve equations (\ref{gs_onepoint_eqs}) numerically, we employ the leap-frog discretization \cite{Borsanyi:0809.4711,Berges:1403.4849} with time step ${\delta t}$, which has the advantage of being numerically stable and phase space volume preserving. We enumerate the discrete steps of the numerical evolution by the discrete variable $\tau = 0, 1, 2, \ldots$. In each evolution step we first update the bosonic momenta:
\begin{eqnarray}
\label{csft_eq_matrix_discrete_P}
 P_i^{\tau+1}
 &=&
 P_i^{\tau} + {\delta t} \lrs{X^{\tau}_j, \lrs{X_i^{\tau}, X_j^{\tau}}}
\nonumber \\
& &-
 i {\delta t} \sum\limits_{\epsilon>0} \sigma_i^{\alpha\beta} \lrs{u_{\alpha}^{\tau}, v_{\beta}^{\tau}},
\end{eqnarray}
where we have used the matrix notation as in (\ref{bfss_Hamiltonian_matrix}) for shortness. Next, the fermionic mode functions are updated:
\begin{eqnarray}
\label{csft_eq_modes_matrix_discrete}
 u_{\alpha}^{\tau+1}
 &=&
 u_{\alpha}^{\tau-1}
 -
 2 i \, {\delta t} \, \sigma_i^{\alpha\beta} \lrs{X_i^{\tau}, u_{\beta}^{\tau}}  ,
 \quad \tau \geq 1,
 \nonumber \\
 u_{\alpha}^1
 &=&
 u_{\alpha}^0 -
 i \, {\delta t} \, \sigma^i_{\alpha\beta} \lrs{X^0_i, u_{\beta}^{0}} .
\end{eqnarray}
Evolution equations for $v_{\alpha}$ take the same form. This evolution scheme corresponds to using the symmetric discretization of the time derivative on the real time axis, which admits also the propagation of lattice doubler modes. Since these doubler modes correspond to Majorana-Weyl fermions of opposite helicity, together with the physical modes they will form $\lr{9+1}$-dimensional Dirac fermions. This enhancement of the fermionic Hilbert space could potentially spoil some of the nice properties of the BFSS model, thus it is very important to prevent the doubler modes from being excited. The second equation in (\ref{csft_eq_modes_matrix_discrete}) ensures that the doubler modes remain practically unexcited for sufficiently long evolution time \cite{Borsanyi:0809.4711}.

After that we update the bosonic correlators $\cev{\hX^a_i \hP^b_j}$ and $\cev{\hP^a_i \hP^b_j}$ according to equations (\ref{gs_twopoint_eqs_XP}):
\begin{eqnarray}
\label{discrete_XP_PP_update}
 \cev{\hX_A \hP_B}_{\tau+1}
 =
 \cev{\hX_A \hP_B}_{\tau}
  + \nonumber \\ +
 {\delta t} \cev{ \hP_A \hP_B}_{\tau}
 + \nonumber \\ +
 3 {\delta t} \cev{\hX_A \hX_E}_{\tau} V_{ECDB} \vev{\hX_C \hX_D}_{\tau}  ,
 \nonumber \\
 \cev{\hP_A \hP_B}_{\tau+1}
 =
 \cev{\hP_A \hP_B}_{\tau}
 + \nonumber \\ +
 3 {\delta t} \, V_{ACDE} \vev{\hX_C \hX_D}_{\tau}
 \times 
 \cev{\hX_E \hP_B}_{\tau+1} ,
\end{eqnarray}
where we have used the short-hand notations of equation (\ref{csft_quartic_2p_eqs}).

Finally, the variables $X^a_i$ and $\cev{\hX^a_i \hX^b_j}$ are updated:
\begin{eqnarray}
\label{discrete_X_update}
 X_i^{\tau+1}
 =
 X_i^{\tau} + {\delta t} \, P_i^{\tau+1},
\\ \label{discrete_XX_update}
 \cev{\hX_A \hX_B}_{\tau+1}
 =
 \cev{\hX_A \hX_B}_{\tau}
 + \nonumber\\ +
 {\delta t} \cev{\hX_A \hP_B}_{\tau+1}
 + 
 {\delta t} \cev{\hP_A \hX_B}_{\tau+1} .
\end{eqnarray}

The commutator representation (\ref{csft_quartic_eqs_matrix}) of the equations (\ref{gs_twopoint_eqs}) allows to devise a better discretization which would involve symmetrized time derivatives, similarly to discrete equations (\ref{csft_eq_modes_matrix_discrete}) for fermionic mode functions. We have not yet implemented this option into our simulations, and simply achieve a comparable precision by using smaller discrete time step ${\delta t}$.

All the data presented in the main text of the paper were obtained with ${\delta t} = 2 \cdot 10^{-5}/\xx$. Such a rescaling with respect to $\xx$ is necessary, since at different $\xx$ the dynamics is characterized by very different time scales which also require different discretization steps to achieve the same accuracy. In order to check the effect of discretization artifacts on our simulations, we have also performed several simulations with two times smaller time step ${\delta t} = 10^{-5}/\sigma$ and checked that these simulations yield practically the same results for all parameter sets which we have used.

\section{Pauli matrices in \texorpdfstring{$d=9$}{d=9} spatial dimensions}
\label{apdx:sigma_identities}

We use the following explicit form of the $\sigma_i$ matrices:
\begin{eqnarray}
 \sigma_1
 &=&
 s_3 \otimes I   \otimes I   \otimes I   \nonumber \\
 \sigma_2
 &=&
 s_2 \otimes s_2 \otimes s_2 \otimes s_2 \nonumber \\
 \sigma_3
 &=&
 s_2 \otimes s_2 \otimes I   \otimes s_1 \nonumber \\
 \sigma_4
 &=&
 s_2 \otimes s_2 \otimes I   \otimes s_3 \nonumber \\
 \sigma_5
 &=&
 s_2 \otimes s_1 \otimes s_2 \otimes I   \nonumber \\
 \sigma_6
 &=&
 s_2 \otimes s_3 \otimes s_2 \otimes I   \nonumber \\
 \sigma_7
 &=&
 s_2 \otimes I   \otimes s_1 \otimes s_2 \nonumber \\
 \sigma_8
 &=&
 s_2 \otimes I   \otimes s_3 \otimes s_2 \nonumber \\
 \sigma_9
 &=&
 s_1 \otimes I   \otimes I   \otimes I   ,
\end{eqnarray}
where $s_1$, $s_2$ and $s_3$ are the conventional $2 \times 2$ Pauli matrices. We note that despite $s_2$ is complex and anti-symmetric, it always enter the $\sigma$ matrices twice. Thus, the $\sigma$ matrices are manifestly real and symmetric.

To demonstrate the conservation of the angular momentum (\ref{angular_momentum}), one also needs the commutation relations between $\sigma$ matrices and their commutators $\sigma_{ij} \equiv \sigma_i \sigma_j - \sigma_j \sigma_i$ (which can be also interpreted as the generators of rotations in the space of Majorana-Weyl spinors):
\begin{eqnarray}
\label{smunu_srho}
 \sigma_{ij} \sigma_k - \sigma_k \sigma_{ij} = 4 \sigma_i \delta_{jk} - 4 \sigma_j \delta_{ik}
\end{eqnarray}

For the proof of the conservation of supersymmetry generators (\ref{bfss_susy_charge}) outlined in Appendix~\ref{apdx:csft_symmetries} one also needs the Fierz identity
\begin{eqnarray}
\label{fierz_identity}
& &
\lrs{\sigma_i}_{\alpha\beta} \lrs{\sigma_i}_{\gamma\delta}
 +
 \lrs{\sigma_i}_{\alpha\gamma} \lrs{\sigma_i}_{\beta\delta}
 +
 \lrs{\sigma_i}_{\alpha\delta} \lrs{\sigma_i}_{\gamma\beta}
\nonumber \\
& &
\qquad\qquad
=
 \delta_{\alpha \beta} \delta_{\gamma\delta}
 +
 \delta_{\alpha\gamma} \delta_{\beta\delta}
 +
 \delta_{\alpha\delta} \delta_{\gamma\beta} ,
\end{eqnarray}
as well as the following identity for the $\sigma_{ij}$ matrices:
\begin{eqnarray}
\label{smunu_double}
& &\sigma_{ij} \sigma_{kl} + \sigma_{jk} \sigma_{li} + \sigma_{kl} \sigma_{ij} + \sigma_{li} \sigma_{jk}
 \nonumber \\
 &&
 \qquad\qquad
 =
 8 \lr{\delta_{ij} \delta_{kl} + \delta_{il} \delta_{jk} - 2 \delta_{ik} \delta_{jl}} .
\end{eqnarray}

\end{document}